\documentclass[titlepage,12pt]{utarticle}   
\usepackage{graphicx}
\usepackage{amsmath,bm,amsfonts,amssymb,amsthm,cite}
\usepackage{hyperref}
\hypersetup{colorlinks=true, linkcolor=black,  anchorcolor=magenta,citecolor=black, filecolor=black, menucolor=black, pagecolor=black, urlcolor=black,bookmarks=true}
\DeclareFontFamily{U}{rsf}{}
\DeclareFontShape{U}{rsf}{m}{n}{<5> <6> rsfs5 <7> <8> <9> rsfs7 <10-> rsfs10}{}
\DeclareMathAlphabet\Scr{U}{rsf}{m}{n}

\newcommand{\al}{\alpha}
\newcommand{\be}{\beta}
\newcommand{\ga}{\gamma}

\newcommand{\e}{\epsilon}

\newcommand{\p}{\pi}

\newcommand {\cC} {{\cal C}}

\newcommand {\cF} {{\cal F}}
\newcommand {\cG} {{\cal G}}
\newcommand {\cH} {{\cal H}}

\newcommand {\cL} {{\cal L}}
\newcommand {\cM} {{\cal M}}

\newcommand {\cQ} {{\cal Q}}

\newcommand {\cV} {{\cal V}}
\newcommand {\cW} {{\cal W}}

\newcommand {\bbC} {\mathbb{C}}

\newcommand {\bbZ} {\mathbb{Z}}

\newcommand{\del}{\partial}

\newcommand{\eg}{{\it eg.}}
\newcommand{\ie}{{\it ie.}}
\newcommand{\Udel}{{U_D}}
\newcommand{\beq}{\begin{equation}}
\newcommand{\eeq}{\end{equation}}
\newcommand{\bea}{\begin{eqnarray}}
\newcommand{\eea}{\end{eqnarray}}
\newcommand{\nn}{\nonumber}
\def\ll{\label}
\newcommand{\beql}[1]{\begin{eqnarray}\label{#1}}
\newcommand{\eeql}{\end{eqnarray}}
\newcommand{\idone}{\bm{1}}
\newcommand{\PSI}[2]{\Psi^{(#1,#2)}}
\newcommand{\ib}{{\bar i}}
\newcommand{\jb}{{\bar j}}
\newcommand{\pb}{{\bar\pi}}
\newcommand{\ft}{{\zeta}}
\newcommand{\rt}{{\rho}}
\newcommand{\eqz}[1]{{(\ref{eq:#1})}}
\newcommand{\lab}[1]{^{(#1)}}
\newcommand{\str}{{\rm str}}
\newcommand{\sdot}{\!\cdot\!}
\newcommand\topa[2]{\genfrac{}{}{0pt}{0}{\textstyle #1}{\textstyle #2}} 
\newcommand{\rhom}{{\rm Hom}^*}

\begin{document}
\preprint{
  CERN-TH-2018-062\\
}

\title{\vskip-1cm On Matrix Factorizations, 
 \\ \vskip.0cm Residue Pairings  and \\ \vskip.3cm  Homological Mirror Symmetry}

\author{
Wolfgang Lerche
} 

 \oneaddress{
 Theoretical Physics Department\\
 CERN, Geneva
 }

    \nobreak
\Abstract{
We argue how boundary $B$-type Landau-Ginzburg models based on
matrix factorizations can be used to compute exact superpotentials
for intersecting $D$-brane configurations on compact Calabi-Yau
spaces. In this paper, we consider the dependence of open-string,
boundary changing correlators on bulk moduli. This determines,
via mirror symmetry, non-trivial disk instanton corrections in the $A$-model. 
As crucial ingredient we propose a differential equation that involves
matrix analogs of Saito's higher residue pairings. 
As example, we compute from this for the elliptic curve certain quantum products  $m_2$ and $m_3$, which
reproduce genuine boundary changing, open Gromov-Witten invariants.}

\date{March 2018}
\maketitle
\tableofcontents
\pagebreak

\section{Introduction}

\subsection{Physical motivation}

Topological open strings in connection with mirror symmetry (for overviews, see \cite{cox1999mirror,mirrbook1,Aspinwall:2009isa}) are useful
for understanding certain non-perturbative phenomena related to
$D$-branes on Calabi-Yau manifolds \cite{Aspinwall:2004jr}, and specifically, for computing exact, 
instanton-corrected effective superpotentials. So far, substantial
progress (initiated in \cite{Aganagic:2000gs,Aganagic:2001nx,Mayr:2001xk})  has been made for single or multiple parallel branes, for which the
open strings are associated with ``boundary preserving'' vertex
operators. Most of these works deal with non-compact  geometries.
There has been much less work on compact geometries (initiated in \cite{Walcher:2006rs,Morrison:2007bm})  and in particular
very little work on intersecting branes, where ``boundary changing''
open string vertex operators come into play and the methods developed
so far are not applicable. 

Such brane configurations are particularly interesting
for phenomenological applications, not the least because they
naturally give rise to chiral fermions; the boundary changing operators
correspond to matter fields in bi-fundamental gauge representations (for a review see
\eg~\cite{Blumenhagen:2005mu}). Such models can be represented by quiver
diagrams, where the nodes correspond to branes and the maps between them
to boundary changing open strings localized at the intersections.
Essentially, the various terms of the effective potential on the world-volume
are given by disk correlators of boundary changing vertex
operators, summed over orderings corresponding to
closed paths in the quiver diagram. 
However, most discussions stop here at the level of cohomology and charge selection rules.

But there is much more to the superpotential than just chasing
arrows around a quiver: generically there are moduli from
the parent Calabi-Yau space and possibly also open string (location and bundle) 
moduli from the branes, and the maps, and consequently the effective potential,  depend
on them.  This dependence can be highly non-trivial
due to infinite series of world-sheet instanton contributions.
Obviously, for answering questions
like what the vacuum structure of the full theory is, one needs to determine
the dependence of the superpotential on these moduli,
over the whole of the moduli space.

\begin{figure}[t]  
\begin{center}
\includegraphics[width=14cm]{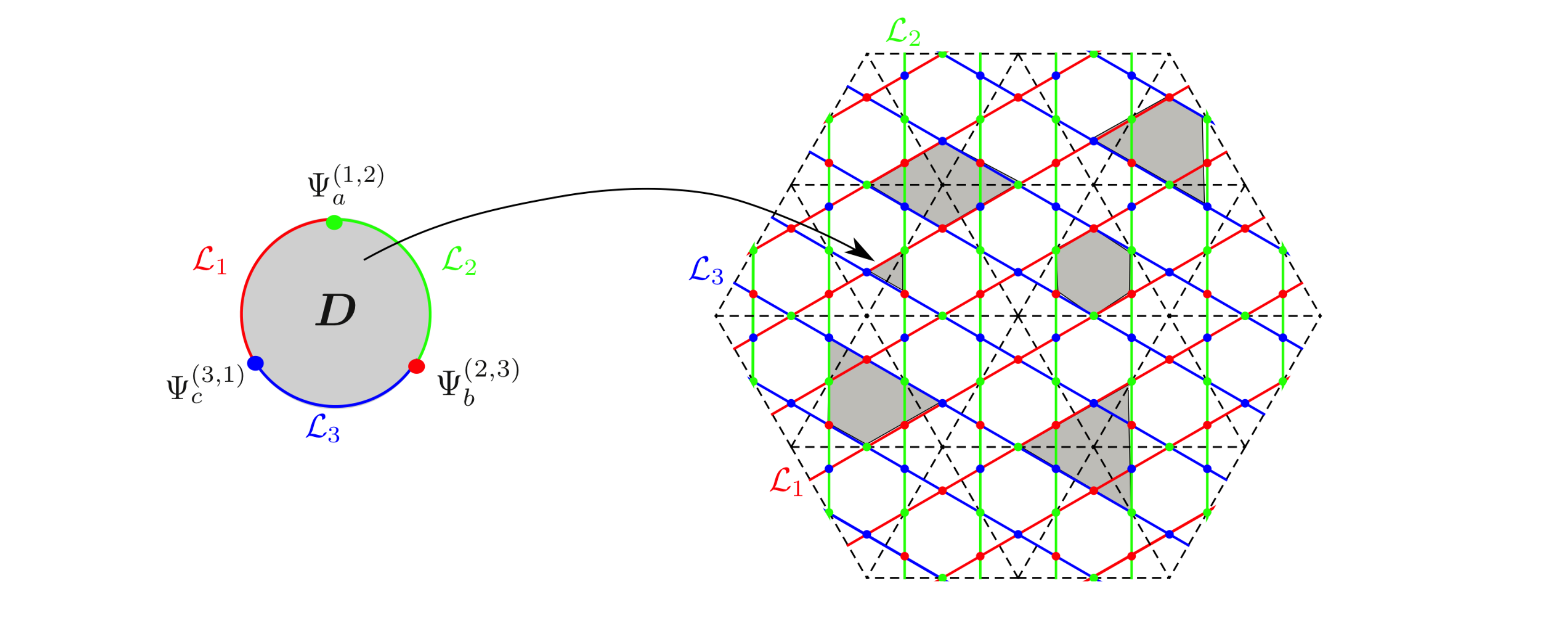}
\caption{In the $A$-model, the boundary changing correlators  get contributions from 
world-sheet instantons that
correspond to holomorphic maps
from the disk world-sheet  into the 
intersecting D-brane geometry (with appropriately matching boundary conditions).
Notice that the instanton problem becomes infinitely rich when we 
allow for intersecting branes, even for the simplest example of the elliptic curve.
}
\ll{fig:instantons}
\end{center}
\end{figure}

Mathematically the instanton problem corresponds to the counting
of holomorphic maps from ``polygon shaped'' disks with $n$ boundary components into Calabi-Yau spaces, such that
these boundaries lie across intersecting special Lagrangian cycles; see Fig.\ \ref{fig:instantons}
for an illustration adapted to the elliptic curve.
There we see that if we allow for generic, $(p,q)$ types of intersecting branes, the set of possible 
correlation functions becomes infinitely much richer as compared
to non-intersecting brane configurations!

The mathematical framework that is designed to address
precisely this kind of questions is homological mirror symmetry \cite{kontsevich,Aspinwall:2009isa}.
However, despite of that this has been an important ongoing topic in mathematics
since more than 20 years, it has seen little use in physics.
Indeed, for example, an explicit method for computing
instanton corrections for intersecting brane configurations 
on Calabi-Yau threefolds folds has been missing so far, even
for the simplest amplitudes such as three-point functions. 

Our purpose in the present paper is to make some modest steps into this direction, 
from an admittedly simple-minded physicist's perspective; we hope that the physics intuition may help
to further the development of the theory.

Let us be more specific. In topological strings, there are two somewhat antagonistic approaches to correlation functions.
One takes a more algebraic, the other a more geometric viewpoint, involving period integrals, the variation of Hodge structures, etc. 
Obviously one would like to have a more unified understanding of both perspectives. 
A sensible first step would be to aim at the middle ground, namely by asking how the algebraic open string sector
varies as fiber over the closed string moduli space. 

This is what we will discuss in this paper,  for open topological strings in the framework of twisted $N=(2,2)$ superconformal field theory.
More specifically, we will consider open string correlation functions on the disk $D$ of the form
\beq
\ll{eq:opencorr}  
B_{\al_0....\al_k}\lab{A_0,..,A_k}(t) =\ 
\Big\langle \Psi_{\al_0} \lab{A_0,A_1}\Psi_{\al_1}\lab{A_1,A_2} P\big(\int_{\del D}\!\!\Psi\lab{1)(A_2,A_3}_{\al_2}\Big)\dots \big(\int_{\del D}
\!\!\Psi\lab{1)(A_{k-1},A_{k}}_{{\al_{k-1}}}\Big)\Psi\lab{A_k,A_0}_{\al_k}
 e^{-\sum t_i\int_D\!\phi_i\lab2}\Big\rangle,
\eeq
which are perturbed by closed string deformations, $\phi$.
Here $\Psi_{*}\lab{A_i,A_j}$ with $i\not=j$
 are vertex operators that describe open strings that go from one boundary condition, or $D$ brane $\cL_{A_i}$, to another one,
$\cL_{A_j}$. Physically they are localized at the intersection $\cL_{A_i}\cap\cL_{A_j}$. These are the
boundary changing operators, in contrast to the
boundary preserving ones,  $\Psi_{*}\lab{A_i,A_i}$, which are localized on one brane only.  The superscripts
denote the integrated $1-$ or $2-$form descendants of the respective operators.

In terms of these correlators, the effective potential that is induced by the $D$-brane background
 is given by summing over all correlators pertaining to the 
given $D$-brane configuration:
\beq
\ll{eq:Weff}
\cW\lab{A_0,..,A_k}_{eff}(s,t) =\ \sum_{k \geq2}{1\over k}s_{\al_0} s_{\al_1} \dots s_{\al_k} \,B\lab{A_0,..,A_k}_{\al_0....\al_k}(t) ,
\eeq
where $s_\al$ are the (not necessarily commuting) fields in the effective action that source the $\Psi_\al$.
This amounts to summing over appropriate closed paths in the quiver diagram.

\subsection{Mathematical setting}

The general structure of open string correlators on the disk is well-known 
(see \eg~\cite{Lazaroiu:2001nm,Kajiura:2003ax}):  they can be written as
\beq
\ll{eq:mopencorr}
B_{\al_0....\al_k}(t)\ =\ \Big\langle\Big\langle\Psi_{\al_0}, m_k( \Psi_{\al_1}\otimes \Psi_{\al_2}\dots \otimes \Psi_{\al_k}) \Big\rangle\Big\rangle\\ ,
\eeq
where the topological metric $\langle\langle *,*\rangle\rangle$ denotes a suitable, non-degenerate and cyclically symmetric inner product.
Moreover, $m_k:\Psi^{\otimes k}\to \Psi$ are certain higher multilinear, non-commutative products that take
collections of operators as input and produce one operator as output.  
Correspondingly, the equations of motion arising from $\cW_{eff}$ take the form 
\beq
\ll{eq:MCs}
\sum_{k\geq0} m_k({\Psi_*}^{\otimes k})\ =\ 0,
\eeq
which are nothing but the Maurer-Cartan equations \cite{Fukaya:2001uc} which specify the locus of unobstructed deformations of the theory.

The products $m_k$, and thus the correlators built from them, satisfy a host of Ward identities, specifically the $A_\infty$ relations,
and have certain cyclicity  and integrability properties.
These data together with the inner product and some other subsidiary conditions comprise what is called a Calabi-Yau $A_\infty$ algebra 
\cite{Costello:2004}.
If there are several boundary components present, the $A_\infty$ algebra is promoted to an $A_\infty$ category.
Moreover, since we have deformed the correlators by bulk moduli, we encounter deformed $m_k=m_k(t)$,  which form what is called a curved $A_\infty$ structure. When combined with the bulk sector, this structure is promoted to an open-closed homotopy algebra \cite{Kajiura:2005sn}.
All this has been discussed at length in the literature,  and we do not want to spend more than a few remarks on this
in  the present context. Suffice it to refer the reader to refs.~\cite{Lazaroiu:2001nm,Kajiura:2003ax,Herbst:2004jp,Carqueville:2011vr}  
for more details, from a physics perspective.

\begin{figure}[t]  
\begin{center}
\includegraphics[width=14cm]{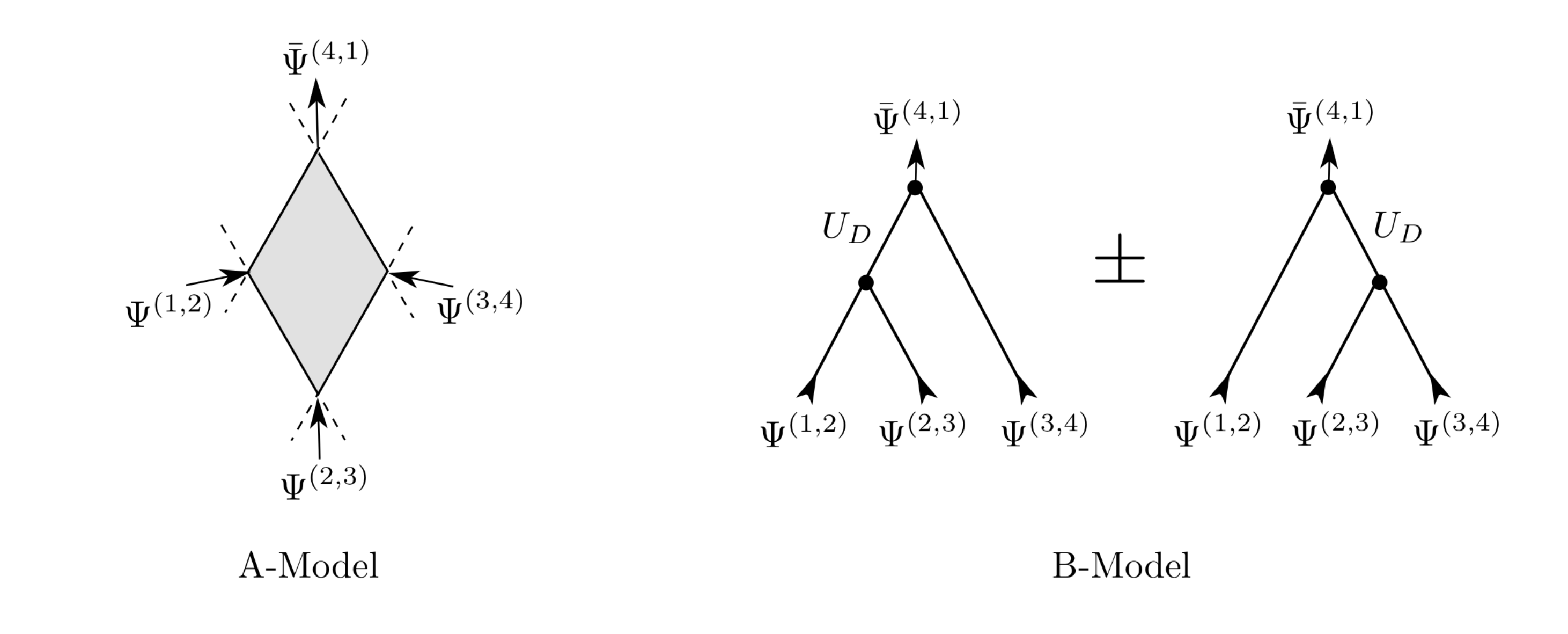}
\caption{Homological mirror symmetry between  $A_\infty$ products $m_3:\Psi^{\otimes3}\to\bar \Psi$.
The value of the Fukaya product in the $A$-model is given by the exponentiated disk instanton action associated with the intersecting brane geometry.
In the $B$-model the corresponding Massey product is computed by nested trees, involving propagators and lower trees. 
}
\ll{fig:HMS}
\end{center}
\end{figure}

In practice, the explicit evaluation of the correlators \eqz{opencorr} is difficult, not the least because of contact terms that arise when the
integrated insertions hit other operators, or $\int\phi\lab2$ hits the boundary. While the  underlying algebraic structure organizes these contact terms, it is not of direct help to actually compute the correlators. As we will see, one needs to augment the algebraic structure by certain differential equations. 

An analogous problem appears already for closed strings, where special geometry in connection with
mirror symmetry comes to the rescue \cite{mirrbook1}. Specifically, mirror symmetry
is the statement that the topological $A$-model on a Calabi-Yau space $Y$ is equivalent to the topological $B$-model on the mirror manifold, $X$. This means that instanton corrected correlation functions in the $A$-model can be computed in terms of classical correlation functions in the simpler $B$-model, in the framework of topological Landau-Ginzburg models.
 For non-intersecting $D$-brane geometries, a corresponding geometrical framework of open mirror symmetry has been well developed
\cite{mirrbook1,Aspinwall:2009isa} after the initial works \cite{Hori:2000ck,Aganagic:2000gs,Aganagic:2001nx,Mayr:2001xk}.

As said before, our intention is to push the subject to more general, intersecting brane configurations, which is the arena of homological mirror symmetry.
Given that this is mathematically highly sophisticated, it would be impossible to give any reasonable account here. Instead we will exhibit
only some basic ideas to convey the motivation of what we want to do.
In short, the basic statement is an isomorphism \cite{kontsevich}:
\beq
\ll{eq:HMS}
Fuk(Y)\ \simeq  D^b(Coh(X))\ ,
\eeq
where $Fuk(Y)$ and $D^b(Coh(X))$ denote the appropriately defined Fukaya category of special lagrangian cycles on the $A$-model side,
and the bounded derived category of coherent sheaves (crudely: vector bundles on submanifolds) on the $B$-model side.
It has been proved recently \cite{Sheridan:2015c} for hypersurfaces in projective space.
The relevance for the physics of $D$-branes has been recognized early on in refs.~\cite{Douglas:2000gi,Lazaroiu:2001jm}.
Equation \eqz{HMS} implies isomorphisms between the $A_\infty$ products $m_k$ on the two sides, 
which are called Fukaya and Massey products
in the $A$- and $B$-model, respectively. See {Figure}~\ref{fig:HMS} for a visual representation.
In physics language this amounts to an ``equality'' of the respective correlation functions. 
We put the word equality in quotation marks, since isomorphism means equality up to maps. 
The physicists are however interested
in explicit expressions, not just in structural existence proofs of isomorphisms. In the closed string sector, this has been achieved
\cite{Candelas:1990rm} by applying methods of algebraic geometry, in particular the theory of Hodge variations, period integrals
and associated flatness differential equations. This leads to an exact map, the mirror map 
\beq
\ll{eq:VSHS}
t(\al): V^B(X)\simeq H^*(X, \wedge^*T_X ) \ \to\  V^A(Y) \simeq QH^*(Y) \ ,
\eeq
which maps the algebraic modulus $\al$ in the $B$-model into the flat coordinate $t$ of the $A$-model. 
Above, $V^B$ and $V^A$ denote the variation of Hodge structures \cite{Barannikov:2000ed,Barannikov2002}
 on the both sides, respectively.
They each comprise of data $V=(H^*,\nabla^{GM},\langle\langle*,*\rangle\rangle)$, where 
$H^*$ is the relevant cohomology or quantum cohomology, $\nabla^{GM}$ the respective Gauss-Manin connection
and $\langle\langle*,*\rangle\rangle$ a suitable inner product, or pairing.

 A corresponding theory has been developed for homological mirror symmetry, which is the theory of non-commutative Hodge variations 
(see \eg \cite{Kontsevich:2006jb,Katzarkov:2008hs,Soibelman:2004}).
The details not being important here, we will be schematic and refer the reader to the readable expositions in
refs.\ \cite{Sheridan:2015a,Sheridan:2015b}.  The basic notion is an $A_\infty$ category $\cC$ with
maps between objects $\cL_i$:
\beq
\ll{eq:Cdef}
\cC_k(\cL_0,\cL_1,\dots \cL_k)\ =\ \rhom(\cL_0,\cL_1)\otimes\rhom(\cL_1,\cL_2)\dots  \otimes\rhom(\cL_{k-1},\cL_k).
\eeq
With this one forms the Hochschild chain complex and its homology
\bea
\ll{eq:HSC}
CC_\bullet(\cC)&=& \bigoplus_k \cC_k(\cL_0,\cL_1,\dots \cL_k,\cL_0),\\
HH_* (\cC)&=& H^*(\cC\cC_\bullet(\cC),b)\ ,\nn
\eea
where $b$ is the Hochschild differential. 
One may refine this to a semi-infinite variant and consider the so-called negative cyclic complex, $CC_\bullet^-(\cC)$, and its cohomology, $HC_\bullet^-(\cC)$.
This involves the introduction of a spectral parameter, $u$.
Together with the Gauss-Manin-Getzler connection $\nabla^{GMG}$ \cite{Getzler93} and the pairing  \cite{Shklyarov:2011},
this forms a structure $\cV(\cC)=(HC_\bullet^-(\cC), \nabla^{GMG}, \langle\langle*,*\rangle\rangle)$, which is a non-commutative analog
of the semi-infinite variation of Hodge structures $V$ in the closed string theory.
In the present context, there are two versions of this, one for the $A$-model side and one for the $B$-model side.  
Homological mirror symmetry then amounts to an isomorphism, in analogy to \eqz{VSHS}:
\beq
\ll{eq:NCVSHS}
\cV^A(Fuk(Y) \simeq \cV^B(D^b(Coh(X))\ .
\eeq
It has been shown \cite{Sheridan:2015a,Sheridan:2015b} that (under suitable conditions) open-closed maps
 exist that provide isomorphisms
\bea
\ll{eq:OCIS}
OC:\qquad\cV^A(HC_\bullet^-(Fuk(Y)))\qquad \ &\to& V^A(QH^*(Y))\\
\cV^B(HC_\bullet^-(D^b(Coh(X))))\ &\to& V^B( H^*(X)).\nn
\eea
This means that the non-commutative Hodge structures of the open string theories map back to the Hodge structures of the bulk theories,
and this can be used to show that homological mirror symmetry implies Hodge theoretical mirror symmetry  \cite{Sheridan:2015a,Sheridan:2015b}. 

What does this tell us for our problem, namely computing the open string correlation functions \eqz{opencorr}? 
These correspond to the components of the Hochschild complex \eqz{HSC} and the statement is that their deformation theory is
isomorphic to the one of the closed string. This by itself does not fix any correlator. 
We would need an analog of the mirror map \eqz{VSHS} to explicitly tie the $A$- and $B$-models together.     
For this, we should impose some extra structure in the form of flatness equations. The question is how to do this in practice.

Flatness equations based on the Gauss-Manin-Getzler\footnote{There are two kinds, one associated with moduli deformations, $t$, plus one associated with the spectral parameter $u$ if we are interested in the semi-infinite extension of Hodge variations; we will suppress the latter, as for our
current purposes this extension is not immediately relevant.} connection
 \cite{Getzler93}, $\nabla^{GMG}_t$,
 are a central theme in the mathematical literature. However, 
the latter has been mostly concerned to map via $OC$ all kinds of quantities to the bulk, closed string sector. 
This allows to evaluate pairings and higher correlators in the simpler, commutative theory.
Specifically, for two-point functions, one might be tempted to write
\beq
\ll{eq:OCopencorr}
\langle\langle\Psi\lab{A,B},\Psi\lab{B,A}\rangle\rangle_D \ =\  \langle\langle OC[\Psi\lab{A,B}],OC[\Psi\lab{B,A}]\rangle\rangle_{S_2},
\eeq
and consider flatness equations of the form
\bea
\ll{eq:intertwine}
0\ =\ \langle\langle \nabla^{GMG}_t  \Psi\lab{A,B},\Psi\lab{B,A}\rangle\rangle_D\   &=&
  \langle\langle OC[\nabla^{GMG}_t  \Psi\lab{A,B}],OC[\Psi\lab{B,A}]\rangle\rangle_{S_2} \\
  &=& \langle\langle \nabla^{GM}_t OC[ \Psi\lab{A,B}],OC[\Psi\lab{B,A}]\rangle\rangle_{S_2}\nn,
 \eea
where the last line exhibits the intertwining property \cite{Sheridan:2015a} of $\nabla^{GMG}_t$ and $\nabla^{GM}_t$ over OC.
However, such correlators may make sense for the annulus, but not for the sphere or disk. Moreover,
the map OC invariably vanishes on boundary changing operators. 
Indeed, $OC$ as well as $\nabla^{GMG}_t$ act on
complete cycles $\Psi\lab{A_0,A_1}\otimes\dots\otimes \Psi\lab{A_k,A_0}$, but not on the individual maps
$\Psi\lab{A_i,A_j}$ when $i\not=j$. 
Thus we cannot capture in this way the boundary changing sector we are interested in.

Rather, the differential operators we seek should act on the individual operators in a well-defined manner, even
if these are boundary changing. This is then finally, in essence, the problem that we want to address: 
find an explicit construction of  flatness differential equations of the form
\beq
\ll{eq:flatparing}
\langle\langle\nabla_t\Psi\lab{A,B},\Psi\lab{B,A}\rangle\rangle_D\ =\ 0 \,,
\eeq
that make sense especially also for boundary changing sectors. 
This will be a crucial ingredient for computing general correlation functions for intersecting branes.

\subsection{Content of the paper}
\ll{intro13}

Our strategy is guided by a specific realization of the open topological $B$-model, namely 
in terms of a $2d$, $N=(2,2)$ superconformal Landau-Ginzburg model based on matrix factorizations. The key point is
an isomorphism
\beq
\ll{eq:DCMF}
D^b(Coh(X))\ \simeq \ Cat(M\!F,W_X)\ ,
\eeq
where $Cat(M\!F,W_X)$ is the category of matrix factorizations of a function $W_X$.  Here this function is given by the LG superpotential
whose vanishing describes the Calabi-Yau manifold $X$ under consideration: $W_X=0$
 (in some weighted projected space). This isomorphism has been proved under certain assumptions, by Orlov \cite{Orlov:2005},
based in earlier ideas of Kontsevich.

For physicists this isomorphism allows to describe topological $B$ type D-branes in terms of a simple field theoretical model,
in which all the elements of the abstract category, namely objects and maps between them, have a concrete realization
in terms of matrix valued field operators. First works on this by physicists include 
\!\!\nocite{Kapustin:2002bi,Brunner:2003dc,Kapustin:2003ga,Kapustin:2003rc}
\cite{Kapustin:2002bi}-\!\!\cite{Kapustin:2003ga},
and a selection of works relevant for our current purposes is given by
\!\!\nocite{Ashok:2004zb,Hori:2004ja,Hori:2004zd,Brunner:2004mt,Walcher:2004tx,Govindarajan:2005im,Govindarajan:2006uy,Knapp:2007kq,Aspinwall:2007cs,Jockers:2007ng,Knapp:2008uw,Carqueville:2009ay,Carqueville:2011px}
\cite{Ashok:2004zb}-\!\!\cite{Carqueville:2011px}.

In the next section we will first briefly review well-known aspects of the Landau-Ginzburg model in the bulk, and study its deformations.
Here we will use a less geometrical and more field theoretical language in terms of renormalization and contact terms.
For this we will introduce the language of Saito's higher residue pairings, 
which so far did not receive a lot of attention in the physics literature. That is why we will use
simple terms, and also provide a sample computation for the elliptic curve. 
It demonstrates how one can formulate differential equations directly in terms of residue pairings.

In Section 3 we will first review some basic facts of matrix factorizations and their deformations. We then analyze
the interplay between bulk-boundary and boundary-boundary contact terms. 
This leads in Section 3.3 to a proposal for a flatness differential equation, which is based on a higher variant of the Kapustin-Li supertrace
residue that plays an open string analog of Saito's first higher residue pairing.

In Section 4 we apply these ideas to open strings on the elliptic curve and demonstrate
 how one can explicitly compute, by solving the differential equation,
open string correlators such as in \eqz{opencorr}. We then match the results of the $B$-model to
the instanton geometry of the $A$-model mirror, thereby reproducing known results. 
Indeed, homological mirror symmetry is very well understood for the elliptic curve (see 
\eg \cite{Polishchuk:1998db,Polishchuk_homologicalmirror,Polishchuk:2000kx,Sato:2017jan}), and
the structure of bundles over it and their sections (given by theta functions), are more or
less explicitly known. Basically this is because the curve is flat and so one can simply read off
the areas of disks using elementary geometry. 

The main purpose of the present paper is not to compute new correlation functions, rather than to develop
 a method how to compute boundary changing, open Gromov-Witten invariants 
directly from the $B$-model from ``first principles'', without relying on prior knowledge about the $A$-model side
(we put quotation marks because we present a proposal, not a proof).
This will be important for later applications, \eg\ to Calabi-Yau threefolds, where the $A$-model correlators are not known beforehand.

\section{Recapitulation of the closed string, bulk theory }
\subsection{Topological Landau-Ginzburg B-model }

The issue is to determine the dependence of the $B$-model $n$-point
correlation functions on the flat complex structure moduli $t_a$ of a Calabi-Yau manifold, $X$.
As per \eqz{VSHS}, via mirror symmetry these coincide with the K\"ahler
moduli of the $A$-model on the mirror manifold, $Y$.  Consider for example
\beq
\ll{eq:threept}
C_{ab c}(t)\ \ =\  \langle\,\phi_a\phi_b\phi_ c e^{\sum t_d\int\phi^{(2)}_d}\rangle\  .
 \eeq
 Here, $\phi_a$ are zero-form cohomology representatives and the deformations 
\beq
\ll{eq:twoform}
\int\!\phi^{(2)}\equiv\int \!d^2z\,\{G^-_{-1},[G^+_{-1},\!\phi]\}\ 
 \eeq
are given by their two-form descendants.
Expanding out we can write
\beq
\ll{eq:texpan}
\langle\,\phi_a\phi_b\phi_ c \int\phi^{(2)}_{d_1}...... \int\phi^{(2)}_{d_n}\rangle\  =: \langle\langle\,\phi_a,
  \ell_{n+2}(\phi_b,\phi_ c,\phi_{d_1},....,\phi_{d_n}\rangle\rangle\ ,
 \eeq
 where the maps $\ell_n$ are the closed string versions of the $m_k$, which obey certain recursive $L_\infty$ relations.  
This statement by itself is not useful to actually compute the correlation functions. 
However, it is here where geometry beats algebra: due to the special geometrical properties of (topologically twisted) $N=(2,2)$ 
superconformal theories,
these correlators satisfy a host of identities  \cite{Witten:1989ig,Dijkgraaf:1990qw,Dijkgraaf:1991dj}. One of those is
that correlators with $n>3$ are obtained by taking  derivatives
of the three-point functions, for example
\beq
\ll{eq:fourpoint}
C_{ab cd}(t)\  =\ {\del\over \del t_d} \langle\,\phi_a\phi_b\phi_c e^{\sum t_e\int\phi^{(2)}_e}\rangle\  .
 \eeq
The property (\ref{eq:fourpoint}) expresses an underlying Frobenius structure \cite{Dubrovin:1994hc,Dubrovin:1998}
and implies crossing-type relations of the form 
\beq
\ll{eq:crossing}
{C_{ab}}^e(t) C_{e cd}(t)\ =\ {C_{a d }}^e(t) C_{e bd }(t)\ .
\eeq
These serve as  integrability condition for the existence of a prepotential $\cF$ such that
\beq
\ll{eq:prepot}
C_{ab c}(t)\ =\ {\del\over\del t_a}{\del\over\del t_b}{\del\over\del t_ c} \cF(t)\ .
\eeq
Note that (\ref{eq:fourpoint}) holds
for a special choice of ``flat'' coordinates $t_a$ of the
moduli space. Otherwise the derivatives would need to be replaced by
covariant derivatives, whose connection pieces encode the contact terms.
In this way, the flatness of coordinates is tied to the cancellation of contact terms.

The important insight of refs.~\cite{Vafa:1990mu,Dijkgraaf:1991dj,CECOTTI1991359,Bershadsky:1993cx,Eguchi:1993ty} and others
was that the exact $n$-point correlators, including all contact
term contributions, can be  computed in an effective
$B$-type, topologically twisted Landau-Ginzburg formulation of the
theory.  It is characterized by a superpotential, $W_0(x)$,  which is a holomorphic, non-degenerate and quasi-homogeneous polynomial
that depends on a number of LG fields, $x_i$, $i=1,...,N$.
These are to be viewed as coordinates of the target Calabi-Yau space $X$
defined by $W_0(x)=0$ in some projective (or weighted projective) space.

We will consider (mini-versal) deformations of the complex structure described by
\beq
  \ll{eq:Wdef}
W(x,t)\ =\ W_0(x) - \sum g_a(t) y_a(x)\ ,
 \eeq
where $y_a(x)$ is some basis of the Jacobi ring, $y_a(x)\in {\rm Jac}(W)\equiv\bbC^N[x]\!/dW$.
 In this language, the flat zero-form operators are represented by polynomials in the LG fields defined by
 \beq
  \ll{eq:flatf}
 \varphi_a(x,t) = {\del\over \del t_a} W(x,t)\ .
 \eeq
  To distinguish the various fields, we denote by $\phi(x,t)$ the marginal operators that
are sourced by the moduli $t$, and by $\varphi(x,t)$ generic elements of the chiral ring 
 that can also include relevant besides marginal ones.
General polynomials are denoted by $\xi$.
In terms of these operators,  the three-point functions then localize in the IR to the 
Grothendieck multi-residue \cite{Vafa:1990mu}, \ie, to\footnote{We have temporarily suppressed a $t$-dependent prefactor that
is needed for the proper normalization. Moreover, the integration contour is a real cycle supported on $|d_iW|=\e$.}
 \beq
    \ll{eq:bulkthree}
C_{ab c}(t)\ =\ {\rm res}_{x=0} {\varphi_a (x,t)\varphi_b (x,t)\varphi_c(x,t)\over d_1W\dots d_NW}\ =\ 
\oint\!{dx \over (dW)^N}\,\,
\varphi_a (x,t)\varphi_b (x,t)\varphi_c(x,t)\ ,
 \eeq
from which all other correlators can be generated.  

The issue thus is to determine the dependence of ${g}_a (t)$ on the flat coordinates.
The most familiar way to determine the flat coordinates is to solve the Gauss-Manin, or Picard-Fuchs system of differential
equations that act on period integrals. This system itself can be systematically 
derived via the variation of Hodge structures. This
procedure \`a la Griffith-Dwork \cite{Griff:1969} is standard since many years (for reviews, see \eg~\cite{cox1999mirror,Morrison:1991cd}), 
so we don't embark on it further.
Suffice it to mention that the core structure consists of the vanishing of the Gauss-Manin connection, $\nabla^{GM}$.
Its primary component takes the form
 \bea
  \ll{eq:GMeq}
\nabla_{t_a}^{GM}(\varphi_b)\ &\equiv&\ {\del\over \del t_a } \varphi_b (x,t) \ -\  U(\phi_a  \varphi_b)  \ =\ 0\ , \qquad
{\rm where} \\ 
  \ll{eq:Udef} U(\, \phi_a  \varphi_b\, )&:\equiv&  d_i\left( \,\phi_a  \varphi_b\,\over d_iW\right)_+\ ,
\eea
and this is what determines the coupling function of $\varphi_a(x,t)$. 
The subscript $+$ denotes a projection: it instructs to
expand the polynomial operator product according to the Hodge decomposition
 \beq
  \ll{eq:Hodge}
\phi_a (x) \varphi_b (x) = {C_{ab}}^c \varphi_c(x) + {p_{ab}}^i(x) d_iW(x)\ ,
\eeq
and then to drop negative powers of $x$; that is, the result is determined by the exact piece of the OPE,
 $U(\phi_a  \varphi_b)=d_i{p_{ab}}^i(x)$. 
 
Mathematically, the $U$ term in (\ref{eq:GMeq}) arises
from integration by parts under the period integral in order to reduce the pole order   
\cite{Griff:1969}.
Physically it is a contact term  \cite{Losev:1992tt}  that arises
from integrating out the massive, exact states in the OPE (\ref{eq:Hodge}).  
Via (\ref{eq:GMeq}), their effect is subsumed in the 
$t$-dependent renormalization of
the coupling functions $g_a(t)$. Schematically:
\beq
 \ll{eq:Linftyren}
\varphi_a(x,t)\ =\
 \varphi^0_a(x) + t \,U(\ell_2(\phi,\varphi^0_a))\vert_{t=0}+ 1/2 \, t^2 \,U(\ell_3(\phi,\phi,\varphi^0_a))\vert_{t=0}+\dots\ .
\eeq
The iterative integrating out amounts to going from an off-shell topological string field theory to an
on-shell ``minimal model" with non-trivial higher $L_\infty$ products $\ell_n$; it is the 
special geometry of the  $N=(2,2)$ superconformal theories that allows to sum all these terms up in one swoop.

\subsection{Higher residue pairings }

We  recapitulate the determination of flat coordinates
by using a method that we find especially suitable for the generalization to the boundary theory.  
This is because it avoids period integrals, which we wouldn't know how to generalize to
matrix valued operators when we will face the generalization to open strings.
It also seems more natural physicswise, as it makes direct contact with the underlying field theory.
This is the theory of Saito's higher residue pairings \cite{Saito:1983,Saito:1983b,Saito:1989,Saito2008FromPF}. 
So far it has been rarely discussed in
physics (see however:  \cite{Blok:1991bi,Losev:1992tt,Losev:1995zc,Losev:1998dv,Li:2018rdd}). 

We will be brief and present only a rough outline. 
For more accurate technical details we refer the reader to the recent expositions \cite{Li:2013kja,Li:2013gza,Li:2015zva}.
The basic objects in Saito theory are the higher residue pairings 
\beq
  \ll{eq:pairing}
K(\varphi_a ,\varphi_b )(u) \equiv  \sum_{\ell=0}^\infty u^\ell K^{(\ell)}(\varphi_a ,\varphi_b ) \ ,
 \eeq
where $u$ is the auxiliary degree 2, spectral parameter that plays an important role
in the variation of semi-infinite Hodge structures \cite{Barannikov:2000ed,Barannikov2002,Kontsevich:2006jb}. 
For us, it has nothing to do with gravitational descendants, rather
 its significance is to provide a grading with respect to the number of iterated contact terms; thus only a finite number
 of powers of $u$ will be relevant. 

The pairings can be derived and represented in various different ways. 
We like to outline a physically motivated, field theoretical derivation which reflects the underlying path integral.
Following the notation of ref.~\cite{Herbst:2004ax}, 
we define the BRST operator related to $B$-type supersymmetry, as well as a deformation of it by $u$:
\bea
\ll{eq:BRSTdef}
\cQ_W\ &=&\ \bar\del + \iota_{dW}\ , \\
\cQ_{W,u} &=& \cQ_W+ u\,\del\ ,
\eea
where:
\beq
\bar\del \equiv \eta^\ib d_\ib\ ,
\qquad 
\del \equiv \del_{\theta_i} d_i\ ,
\qquad
 \iota_{dW}\equiv d_iW\del_{\theta_i} 
 \ .
\eeq
Here, $\bar\eta^{{\bar i}}\sim dz^{{\bar i}}\in T^*_X$ and $\theta_i\sim \del/\del{z_i}\in T_X$ 
denote complex fermions that  represent differential forms on $X$ in the usual manner.
Moreover, the Hamiltonian is
\beq
\ll{eq:Hdef}
\cH_\lambda\ \equiv\ \triangle_{\bar\del}+ \ell_W+ \lambda(L_0+ u\,\del \, \overline{\iota_{dW}})\ ,
\eeq
where $\triangle_{\bar\del}=[\,\bar\del,\bar\del^\dagger]$ is the laplacian,
$\ell_W=d_id_jW\del_{\eta_i}\del_{\theta_j}$, and 
\beq
\ll{eq:LKdef}
L_0\ =\ d_\ib d_\jb \bar W(\phi)\eta^\ib\theta^\jb + ||d_i W(\phi)||^2\ 
\eeq
is the zero mode lagrangian. Note that 
\bea
\cH_\lambda&=& \big[\,\cQ_{W,u},\,\cG_\lambda\,\big]\ ,\   {\rm where}  \\
& &\!\cG_\lambda\ =\ \bar\del^\dagger+\lambda\,\overline{\iota_{dW}}\  \\
& &\bar\del^\dagger\equiv d_i\del_{\eta_{\bar i}}\ ,\qquad
         \overline{\iota_{dW}}\equiv \overline{d_iW}\,\theta_i\ .
\eea
After the reduction to the zero modes of $\triangle_{\bar\del}$ in the path integral, the expression for the one-point function, 
$\int e^{-\cH_\lambda}\xi$, of some arbitrary polynomial $\xi$ takes the form:
\beq
\langle\,\xi(x)\, \rangle \ =\ \int dxd\eta d\theta\,\xi(x)\, e^{-\lambda(L_0+ u \del\,\overline{\iota_{dW}})}\ .
\eeq
Since the Hamiltonian is BRST exact, the correlators are topological and (on-shell) independent of $\lambda$.
For  $\lambda\to\infty$, the second term in $L_0$ localizes the integral on
the critical points of $W$ and contributes $\lambda^{-N}|H|^{-2}$,
while the first term yields a factor of the Hessian $\bar H$ of
$\bar W$. This also comes with a factor $\lambda^{N}$ so that the $N$ zero modes of each of $\theta$ and $\eta$ 
of this term dominate the path integral,
and this implies that the potential $\theta$ and $\eta$ dependence of any other, non-exponential term drops at 
$\lambda\to\infty$. 
The sum over the critical points over the remaining $H^{-1}$ can then be converted,
via a standard argument, to the usual residue formula. 
Thus we get 
\beq
\langle\,\xi(x)\, \rangle(u) \ =\ 
\oint\! {dx\over (dW)^N} \,\sum u^\ell(\del h)^\ell \xi(x)\, \ 
\eeq
as perturbation expansion in terms of the propagator
$$
\del h\ \equiv\ \del{\overline{\iota_{dW}}\over ||dW||^2}\ =\ d_i{\overline{d_iW}\over ||dW||^2}\ .
$$
Considering Morse coordinates near every critical point $x^0_i$
by writing $W\to W(x^0_i)+1/2 (x_i)^2$, and then taking the limit ${x_i}^*\to 0$  {\it\`a
l'H\^opital}, we can effectively cancel out the non-holomorphic pieces in $h$.\footnote{One needs to assume isolated critical points for this argument, and as usual this can be achieved by temporarily resolving the singularity by a small perturbation and invoking continuity.  
A more rigorous derivation proceeds by introducing a compact support and cutoff functions as explained in detail in \cite{Li:2013kja}. }
All-in-all we arrive at 
\bea
\langle\,\xi(x)\, \rangle \ &=&\ \oint\! {dx \over (dW)^N}\,L(u,\xi)\ , \ \ \ {\rm where}\\
L(u,\xi)\ &=&\ \sum u^\ell\left(d_i{1\over d_iW}\right)^\ell \xi(x)\, .
\eea
This then gives the following representation of the higher residue pairings
\beq
 \ll{eq:KIOF}
K(\xi_a ,\xi_b )(u)\ =\ 
\oint\! {dx \over (dW)^N}L(u,\xi_a)\, L(-u,\xi_b)\, .
\eeq
The first, constant term in the expansion in $u$: 
\beq
  \ll{eq:K0}
 K^{(0)}(\varphi_a ,\varphi_b )\ =\ \oint\! dx\,{ \varphi_a(x)\varphi_b(x) \over (dW)^N}\ =\ \eta_{ab}\ ,
\eeq
is just the topological two-point function, or metric in field space. The next term can be written in the form:
\beq
  \ll{eq:K1}
 K^{(1)}(\xi_a ,\xi_b )\ =\ {1\over2} \,
\oint\! {dx\over (dW)^N}  \,   \sum_j  {( d_j\xi_a(x))\xi_b(x) -   \xi_a(x)d_j(\xi_b(x)) \over d_jW(x)}\ .
 \eeq
We call this the ``integrated operator form" of the residue pairings.
We also introduce a ``contact term form" of the pairings by writing
 \bea
  \ll{eq:KCT}
K_C(\xi_a ,\xi_b )(u) &=& K^{(0)}( L_C(u,\xi_a),\,L_C(-u,\xi_a)\,),\  {\rm where}\\
L_C(u,\xi) &=&\sum u^\ell    \overbrace{U(U(... U(}^\ell\xi(x)..)), 
 \eea
which is supposedly\footnote{We are unaware of literature where this has been addressed explicitly, except for $\ell=1$.}
 equivalent to the ``integrated operator form",  (\ref{eq:KIOF}).
The importance of this equality for $\ell=1$,
  \beq
  \ll{eq:Lossev}
K_C^{(1)}(\xi_a ,\xi_b )\ =\ K^{(1)}(\xi_a ,\xi_b )\ ,
 \eeq
has been emphasized by Losev \cite{Losev:1992tt}.

As will become clear later, the reason for this naming is that integrated operator insertions have structurally the form
$\int\xi^{(2)}\sim d\xi /dW$, while contact terms involve a projection to polynomials, $U(\xi)=d(\xi/dW)_+$,
and so can be directly translated to the renormalization of operators.

\subsection{Flatness equations {\it sans p\'eriodes}}

We now turn to the determination of flat coordinates and field bases in terms of higher residue pairings,
reviewing only aspects that are of immediate interest for our purposes.

To start, we preliminarily adopt an overall normalization defined by
\beq
\ll{eq:Hess}
K^{(0)}(H,1)\ =\ 1\ ,
\eeq
where $H=\, $det$\,d_id_jW$ denotes the Hessian of the potential $W$ (which is in general not a flat cohomology representative).
As is well known, the physical correlators involving flat cohomology representatives will need to have an
extra rescaling factor, which is closely related to Saito's primitive form and is essentially given by the fundamental period
$\omega_0$ of the Calabi-Yau $X$. 
 
Next we introduce the notion of a ``good basis" \cite{Saito:1983b,Saito:1989} of field operators $\{\varphi_a\}$,  
which is defined by
 \bea
 K^{(0)}(\varphi_a, \varphi_b) &\equiv& \eta_{ab}\ =\ {\rm const.}\ , \nn\\
  \ll{eq:goodbasis}
 K^{(\ell>0)}(\varphi_a, \varphi_b)&=& 0\, .
 \eea
 These equations are not quite sufficient for defining flat bases and coordinates.
 Consider the primary one:
 \bea
  \ll{eq:constmetric}
 0& =& \del_t   K^{(0)}(\varphi_a, \varphi_b)\\
 & =& \ K^{(0)}(\del_t\varphi_a, \varphi_b)+K^{(0)}(\varphi_a, \del_t\varphi_b)-
\oint\! dx\,  {\varphi_a(x) \varphi_b(x)\over (dW)^N} \sum {d_i\phi\over d_iW}\nn \\
  & =& \left(K^{(0)}(\del_t\varphi_a, \varphi_b)-K^{(1)}(\phi\varphi_a, \varphi_b)\right)
  + \left(K^{(0)}(\varphi_a, \del_t\varphi_b)+K^{(1)}(\varphi_a, \phi\varphi_b)\right)\ .\nn
 \eea
The peculiar term on the RHS represents in LG language the insertion of the integrated 2-form operator 
 (\ref{eq:twoform}):
 \beq
  \ll{eq:dpdw}
 \int\!\phi^{(2)}\ \longleftrightarrow\ {d\phi\over dW}\ ,
 \eeq
 and for this reason we called  (\ref{eq:K1}) the integrated operator form of the residue pairing.
 By the identity (\ref{eq:Lossev})  we can rewrite it in terms of the contact term form
  \beq
   \ll{eq:ccform}
0\ =\   K^{(0)}(\del_t\varphi_a-U(\phi\varphi_a), \varphi_b)+
K^{(0)}(\varphi_a, \del_t\varphi_b-U(\phi\varphi_b))\ ,
\eeq
which reproduces the geometrical equation (\ref{eq:GMeq}) (we assume the gauge $U(\varphi_*)=0$ here).
Either way, the equation can be compactly rewritten as
 \beq    
 0 =K(\nabla_t^{GM}\varphi_a, \varphi_b)+K(\varphi_a, \nabla_t^{GM}\varphi_b)\ , 
 \eeq
where
 \beq
 \nabla_t ^{GM} =  \del_t- {\phi\over u}\ .
  \eeq
For complete flatness, requiring the constancy of the topological metric,
eq.\ (\ref{eq:constmetric}) is not sufficient. Rather one needs to
impose the stronger "chiral square root" of this equation, and this for all higher pairings as well:
  \beq
    \ll{eq:fullflatness}
K\lab\ell(\nabla_t^{GM}\varphi_a, \varphi_*)\ =\ 0\ .
\eeq
For any given $\varphi_a$, this equation must hold for 
arbitrary cohomology elements  $\varphi_*$ in a ``good'' basis (\ref{eq:goodbasis}), and it can give non-trivial
constraints if the degrees of the $\varphi_*$ are matched appropriately to the one of $\varphi_a$, and to $\ell$.

Equation \eqz{fullflatness} is in disguise what the first order form of the familiar Picard-Fuchs system instructs us to do.
The essence of the story is this: by scanning over $\ell$  and all possible ``spectators'' $\varphi_*$, we sample all components of
the Gauss-Manin connection. In general, there are higher order contact terms beyond what we wrote in (\ref{eq:GMeq}). In the
more familiar language of period integrals,  these arise from multiple, iterated partial integrations, which reflect the structure of the Hodge
filtration. These are encoded by the nested propagators $U$ \eqz{Udef},
where the order of the nesting is measured by $\ell$. 
Physically, testing the differential equation (\ref{eq:fullflatness}) against all
physical operators samples all possible contact terms.  In conjunction with (\ref{eq:goodbasis}), 
it determines the renormalized coupling functions $g_a(t)$.

\subsection{Example: the elliptic curve}

Let us illustrate the method of residue pairings by an easy example computation.
We will reproduce results that are known since long and have been discussed in the physics literature  in \eg~
\cite{Verlinde:1991ci,Lerche:1991wm,Klemm:1991vw,Eguchi:1993ty}.

We consider the simplest possible case, namely the cubic elliptic curve. 
It is defined as the hypersurface $W=0$ in $CP^2$ where
\beq
\ll{eq:Well3}
W(x,t)\ =\ \ft(t)\left[{1\over 3}\left( {x_1}^3+ {x_2}^3+{x_3}^3\right)- \alpha(t) x_1x_2x_3\right]\ .
\eeq
Here $\alpha(t)$ is the complex structure modulus whose dependence on the
flat coordinate $t$ is to be determined. With hindsight we have also performed an
overall rescaling by a function $\ft(t)$ which needs to be determined as well.
The marginal operator thus is
\beq
\ll{eq:phidef}
\phi(x,t)\ =\del_t\,W(x,t)\ = \ - \ft(t) a'(t) x_1x_2x_3 + { \ft'(t)\over 3 \ft(t)} \sum x_i d_i W(x,t)\ ,
\eeq
where we exhibited that it has an exact piece.\footnote{Such exact pieces correspond to linear combinations
of periods in the more familiar geometrical language.} We are also interested
in a perturbation by
\goodbreak
 the following relevant operators:
\bea
\ll{eq:varphidef}
\varphi_1(x,t)&=&\ft(t) g_1(t) x_1\ , \\
\varphi_2(x,t)&=&\ft(t) g_2(t) x_2 x_3\ ,
\eea
and the main task will be to compute $t$-dependent correlation functions of those.
This requires first of all to determine the functions $\al(t),\ft(t),g_a(t)$.

We have already used as ansatz an initial good basis of operators. Indeed all the operators obey 
eq.\! (\ref{eq:goodbasis}), and in particular we have
\beq
\ll{eq:phizero}
K^{(1)}(\phi,\phi)\ =\ 0\ ,
\eeq
despite that $\phi$ has a non-vanishing exact piece. This extra freedom is allowed due to the antisymmetry of
$K^{(1)}$.

We now in turn impose the various flatness equations. First the most trivial one:
\beq
\ll{eq:oneflat}
 K(\nabla_t^{GM} 1,\phi)\ \equiv\  K^{(0)}(\del_t1,\phi) -K^{(1)}(\phi,\phi)\ =\ 0\ .
 \eeq
 This shows that (\ref{eq:phizero}) is indeed a relevant property.  Next, imposing
 \beq
\ll{eq:phiflata}
 K^{(0)}(\nabla_t^{GM}\phi,1)\ \equiv\   K^{(0)}(\del_t\phi,1) -K^{(1)}(\phi\cdot \phi,1)\ =\ 0\ ,
 \eeq
yields the following differential equation:
\beq
\ll{eq:eqphiflata}
{\ft'(t)\over \ft(t) } \ =\ {\al''(t)\over 2 \al'(t) } - {3 \al(t)^2 \al'(t)\over 2 \Delta}\ ,
 \eeq
 where $\Delta\equiv \al^3-1$ is the discriminant of the curve. Up to a constant, it is solved by
 \beq
\ll{eq:solphiflata}
\ft(t)\ =\ -i\sqrt{{\al'(t)\over 3\Delta}}\  . 
\eeq
The rescaling of $W$ indeed coincides with the fundamental period of the elliptic curve:
 \beq
\ll{eq:fundper}
\ft(t)\ =\ \varpi_0(\al) \equiv\  {1\over 3\al}\, {}_2F_{1}(1/3,2/3,1; 1/\al^3)\ ,
\eeq
as expected on general grounds. It also relates to Saito's idea of a primitive form, which
in this context boils down to a rescaling by $\varpi_0$ \cite{Saito:1983b,Saito:1989,Li:2013kja}.

Moreover, we consider the equation at one level up: 
 \beq
\ll{eq:phiflatb}
 K^{(1)}(\nabla_t^{GM}\phi,\phi)\ \equiv\   K^{(1)}(\del_t\phi,\phi) -K^{(2)}(\phi\cdot \phi,\phi)\ =\ 0\ 
 \eeq
 (all higher ones being empty).
This yields the following non-linear differential equation:
 \beq
\ll{eq:phiflatb1}
\big\{\,z;\,t\,\big\}\ =\ -\,\frac{ 9-8z+8 z^2}{18z^2\Delta^2}\,{z'}^2\ ,
\eeq
where $z\equiv 1/\al^3$ and $\{z;t\}\equiv {z'''\over
z'}-\frac32({z''\over z'})^2$ is the Schwarzian derivative. Via standard arguments
the solution of this equation is given by:
$$
{{3\, \al\, (\al^3 +8) \over  \Delta}~=~  j(t)^{1/3} \,,}
$$
where $j(t)  = q^{-1} + 744 + \dots$ is the familiar
modular invariant function in terms of $q=e^{2\pi it}$. This relation
identifies $\al(t)$ with the Hauptmodul of the modular subgroup $\Gamma(3)$. 
Its inverse coincides with the well-known mirror map of the elliptic curve, given by the ratio of its periods:
$t(\al)=\varpi_1(\al)/\varpi_0(\al)=i/\!\sqrt3\,{}_2F_1(1/3,2/3,1;1-1/\al^3)/ {}_2F_1(1/3,2/3,1;1/\al^3) $.

Finally, we turn to the relevant operators. The equations to consider are
 \bea
\ll{eq:releflat}
 K^{(0)}(\nabla_t^{GM}\varphi_1,\varphi_2)\ &\equiv&   K^{(0)}(\del_t\varphi_1,\varphi_2) -K^{(1)}(\phi\cdot\varphi_1,\varphi_2)\ =\ 0\ ,\\
  K^{(0)}(\varphi_1,\nabla_t^{GM}\varphi_2)\ &\equiv&   K^{(0)}(\varphi_1,\del_t\varphi_2) +K^{(1)}(\varphi_1,\phi\cdot\varphi_2)\ =\ 0\ ,
 \eea
which give rise to $g'_a/g_a=a[\ft'/3\ft-\al^2\al'/\Delta]$. The solutions are:
 \beq
\ll{eq:solflat}
g_a(t) \ =\ (\Delta(t) \ft(t))^{a/3}\ ,   \qquad a=1,2.
\eeq
We can similarly determine contact terms between the relevant operators, by defining them
directly in terms of residue pairings, \ie,
\beq
\ll{eq:CTK}
CT[\varphi_a,\varphi_b]\ :=\ \sum_*K^{(1)}((\varphi_a\cdot \varphi_b,\varphi_*)\,\varphi_{\bar*}\ ,
\eeq
where $\varphi_{\bar*}$ is the dual of $\varphi_*$ with regard to the inner product defined by $K^{(0)}$.
Explicitly:
\bea
CT[\varphi_1,\varphi_1] &=& 0\ ,\\
CT[\varphi_1,\varphi_2] &=&    -{\ft'\over \ft^2\Delta}g_1 g_2\cdot 1 \ ,\\
CT[\varphi_2,\varphi_2] &=&   - {\al\over\ft\Delta^2} {g_2}^3 \cdot \varphi_1  \ .
\eea
Usually these terms are included in the LG potential as higher order corrections, which leads to a redefinition of the flat fields.
We don't need to present these formulas here, the interested reader may consult refs.~\cite{Verlinde:1991ci,Lerche:1991wm,Klemm:1991vw}
 for details.

Now we are in the position to determine correlation functions. First note that because of the flattening rescaling 
of $W$ by $\ft$, we need to put a normalization factor to the inner product such as to restore (\ref{eq:Hess}), \ie~$\langle H\rangle=1$:
\beq
\ll{eq:normaliz}
\langle\langle\, \xi_a,\xi_b \rangle\rangle := {1\over \ft^3} K^{(0)}(\xi_a,\xi_b)\ .
\eeq
This is the natural normalization in the current setup, but
might seem peculiar since usually correlators are rescaled by ${1/{\varpi_0}^2}$.
However all is fine because the operators have been consistently rescaled as well.

We now simply plug the renormalized operators in and find:
\bea
\ll{eq:simplestcorr}
\langle\langle\, \phi, 1 \rangle\rangle(t) \ =\ \langle\langle\, \varphi_1, \varphi_2 \rangle\rangle(t) &=& 1\ ,\\
\langle\, \varphi_1 \varphi_1 \varphi_1\rangle(t)  =\  \langle\langle \varphi_1, \varphi_1 \varphi_1  \rangle\rangle(t)  &=& \ft(t) \ ,\\
\langle\, \varphi_1 \varphi_{1,2}\varphi_{1,3}\rangle(t)  =\  \langle\langle \varphi_1, \varphi_{1,2} \varphi_{1,3}  \rangle\rangle(t)  &=& \al(t)\ft(t) \ ,
\eea
(where $\varphi_{1,i}=\varphi_1\vert_{x_1\rightarrow x_i}$).
As an example for a higher point function,  consider
\bea
\ll{eq:fourptcorr1}
\langle\, \varphi_1 \varphi_1 \varphi_2\int \varphi_2^{(2)}\rangle(t)&=&
\langle\langle\, \varphi_1, \ell_3(\varphi_1, \varphi_2,\varphi_2)\rangle\rangle\ 
\ ,\ \ {\rm where}\\
\ell_3(\varphi_1, \varphi_2,\varphi_2)&=& 
2\,CT[\varphi_1, \varphi_2] \cdot \varphi_2 +    CT[\varphi_2,\varphi_2]\cdot  \varphi_1     \\
&=& -{1\over \ft\Delta}\left[ 2 {\ft'\over \ft}g_1 g_2\,\varphi_2+ {\al\over\Delta} {g_2}^3 \,\varphi_1\cdot\varphi_1\right]\ .
\eea
The end result is then
\beq
\ll{eq:fourptcorrres}
\langle\, \varphi_1 \varphi_1 \varphi_2\int \varphi_2^{(2)}\rangle(t)\ =\ {2 \al^2 \al'\over \Delta}-{\al''\over \al'}(t)\ .
\eeq
All of these correlators reproduce results that are known in the literature \cite{Verlinde:1991ci,Lerche:1991wm,Klemm:1991vw,Eguchi:1993ty,Satake2011}.
Our purpose was to re-derive them in terms of a more
field theoretical and less geometrical language that can be easier generalized to open strings.

\section{Open String $B$-Model }

\subsection{Matrix factorizations}

We like to generalize the considerations of the previous sections
to the open string sector, \ie, to the $B$-model on the disk. As mentioned in Section \ref{intro13},
for the relevant boundary LG model the various possible $B$-type topological $D$-branes
are one-to-one to the various possible matrix factorizations $\cM(Q)$ of the
bulk superpotential:
\beq
\ll{eq:Qsquare}
\cM(Q):\ \ Q(x,t)\cdot Q(x,t)\ =\  W(x,t)\,\idone_{n\times n}\ .
\eeq
Here $Q(x,t)$ is the boundary BRST operator that can be represented by
an odd $n\times n$ dimensional matrix (so $n$ is even), whose precise structure
encodes the brane geometry we want to describe. This includes 
also a specific dependence of closed ($t$) and possibly open ($u$) string moduli.
The dimension $n$ of the Chan-Paton space can take arbitrary even values, and this reflects that 
there is in general an infinite number of possible (potentially reducible) brane configurations
on a given CY space, $X$. When $n=2^k$ for some integer
$k$, one can write $Q$ compactly in terms of boundary fermions $\pi,\pb$ that form a Clifford algebra.
This is what we will do, while not being essential.

In the following we will focus on the most canonical of such matrix factorizations,
for which we make use of the quasi-homogeneity of the superpotential,
$W(x,t)=1/2 \sum q_i x_i d_i W(x,t)$. Here the BRST operator is given by
\beq
\ll{eq:Qcan}
Q(x,t)\ =\ 1/2\,q_i\pi_i x_i + \pb_i d_iW(x,t)\ , \ \ \  \{\pi_i,\bar\pi_j\}=\delta_{ij},\  i,j=1...N,
\eeq
where $q_i$ are the $R$-charges of the LG fields.\footnote{For simplicity, we put them all
equal to each other, $q_i=2/N$.  They can be reinstated easily for covering weighted projected spaces as well.} 

The physical operators at the boundary are then given by the non-trivial cohomology classes of $Q$.  
A new feature as compared to the bulk LG theory is that they are
matrix valued. Moreover in general there are boundary preserving and boundary changing operators.
Boundary preserving operators, denoted by $\Psi^{(A)}\equiv\Psi^{(A,A)}$, are each tied to
a single matrix factorization $\cM(Q^{(A)})$, and are represented  by $n_A\times n_A$ matrices.
Boundary changing operators $\Psi^{(A,B)}\in Hom^*(\cM(Q\lab A),\cM(Q\lab B))$ are 
associated with pairs of matrix factorizations, $Q^{(A)}$ and $Q^{(B)}$.
They are represented by not necessarily quadratic, $n_A\!\times\! n_B$ dimensional matrices.  

For a given pair of boundary conditions $(*,*)$, the (``on-shell") space of physical operators
is defined by
\beq
\cH_P^{(*,*)}\ =\  \left\{\PSI **:  \left[ Q,\PSI **\right]=0,\, \PSI **\not=\left[ Q, ....\right]\right\},
\eeq
where\footnote{We
will always denote by $[-,- ]$ the graded commutator which
acts by definition on even and odd elements with the proper sign, and
when boundary changing operators are involved, it implicitly acts from left and
right in the manner defined here.}
\beq
\ll{eq:Qcomm}
\left[\,Q\,,\Psi^{(A,B)}\,\right]\ 
:\equiv\ Q^{(A)}_{n_A\times n_A}\Psi^{(A,B)}_{n_A\times n_B}-
(-1)^s\Psi^{(A,B)}_{n_A\times n_B}Q^{(B)}_{n_B\times n_B}\ ,
\eeq
with $s=0$ or $s=1$ depending on the statistics of $\Psi$.
 
Another difference as compared to the bulk theory,  where variations of cohomology elements
map back to cohomology elements, is that in the boundary theory variations of matrix valued cohomology elements
are in general not BRST invariant:  $[Q(t),\del_t\Psi(t)]\not=0$.
In other words, variations of such cohomology elements will generically map into the full, off-shell Hilbert space at the
boundary. Thus we need to pay attention to the structure of the full Hilbert space, which is given by general, $\bbZ_2$ graded
matrices $M$ with polynomial entries in $x$.
By a basic theorem of Hodge and Kodaira, the latter can
always be decomposed as follows:
\beq
\ll{eq:cohsplit}
\cH\ =\ \cH_{U}\oplus \cH_{P}\oplus\cH_{E}\ .
\eeq
Here $\cH_{U}$ comprises the set of unphysical operators that are not
annihilated by $Q$ and $\cH_{E}$ comprises the exact operators that are
$Q$-variations.  Note that {\it a priori} $\cH_{P}$ is defined only up to
addition of operators in $\cH_{E}$, and similarly $\cH_{U}$
is defined only up to addition of operators in both $\cH_{P}$
and $\cH_{E}$. A given decomposition is referred to
as a cohomological splitting, and can be viewed as a gauge choice
for the off-shell physics (see \eg~\cite{Lazaroiu:2001nm,Kajiura:2003ax}). 

This is closely related to the choice of cohomology representatives. We have seen above that
for the bulk theory, the proper choice of operators including definite exact pieces is an important ingredient
in the determination of flat coordinates. Thus the question arises as to what preferred basis of
operators to choose initially.

The choice of cohomological splitting corresponds to how the inverse of the
BRST operator $Q$ is precisely defined. This inverse, or ``homotopy'' will also
be needed for the computation of higher point correlation
functions, in the form of the open string propagator $\Udel$ that enters in the higher $A_\infty$ products.
Its choice corresponds to the choice of an off-shell completion of the theory, 
and mathematically speaking corresponds to adopting a specific choice of
the minimal model of the underlying $A_\infty$ algebra. 

A good strategy \cite{Sheridan:2015c} to construct the inverse of $Q$ is to regard the boundary sector as fundamental and
the bulk sector as a perturbation by the superpotential, $W$.  That is, we split
\beq
\ll{eq:Qsplit}
Q\ =\ Q_S+Q_W\,, \ \ Q_S\ =\ 1/N\,x_i\pi_i\,,\ \ Q_W=\pb_id_iW\ ,
\eeq
and first invert $Q_S$. We must require that the inversion works properly
on the full Hilbert space, \ie, on arbitrary matrices $M(x)\in\cH$ with polynomial entries. 
To this end, let us make an ansatz 
\beq
U_S\ =\ \kappa^{-1}(M)\,\pb_i d_i\ ,
\eeq
and introduce projection operators
$\Pi_*:\,\cH\rightarrow \cH_*$ with:
  \begin{eqnarray}
    \ll{eq:PI}
\Pi_{E}&=& Q_S \cdot U_S\nn\\
\Pi_{U}&=& U_S \cdot Q_S\\
\Pi_{P}&=& 1-\Pi_{E}^S-\Pi_{U}^S\ ,\nn
 \end{eqnarray}
 where $Q_S$ acts as graded commutator as in \eqz{Qcomm}.
We need to determine the coefficients $\kappa(M)$ such that  these operators are indeed good projectors that
satisfy $(\Pi_\ast)^2\sdot M=\Pi_\ast\sdot M$. They depend on the concrete matrices on which $U_S$ acts, 
and for determining them, we need
to adopt some normal ordered form for $M$ to map back to. For example,
\beq
M\ =\ {\rm const}(a,b,c)\,\pb^a\pi^b x^c\ ,
\eeq
where $a,b,c$ are multi-indices labeling all of $\pb_i,\pi_j,x_k$. The condition of good projections then gives
$\kappa(M)=\sum (a_i+c_i)$.

 The full $\Udel$ that inverts the complete boundary BRST operator $Q$ can then be obtained by a
 simple application of homological perturbation theory, \ie,
\beq
\ll{eq:UfullI}
\Udel\ :=\ U_S\cdot \sum (Q_W\cdot U_S)^l\ .
\eeq
The sum terminates at a finite number of steps and so yields an exact result.
By construction, $Q$ and $\Udel$ satisfy $(\Pi_\ast)^2=\Pi_\ast$ when used in the projectors \eqz{PI},
and so $\Udel$ is indeed a good propagator in the full theory with non-vanishing superpotential $W$.

Note that this construction is not unique: we may equally well define 
an action of $\Udel$ from the right and consider a different normal ordering:
$M={\rm const}(a,b,c)\,\pi^a\pb^b x^c$; this gives $\kappa(M)=\sum (b_i+c_i)$.
In general we may consider linear combinations which also invert $Q$:
\beq
\ll{eq:Ugen}
\Udel(\e)\ :=\ {1\over2}(1+\e)\Udel^L+ {1\over2}(1-\e)\Udel^R\ ,
\eeq
where $\e$ parametrizes the normal ordering ambiguity which leads to an ambiguity of the cohomological splitting of $\cH$
induced by the projectors $\Pi_*(\e)$. Fortunately, the ambiguity cancels when acting on BRST closed operators in $\cH_P\oplus \cH_E$, and
won't play a role in the following. However, we note as a side remark that 
\beq
\Udel(\e)\cdot Q\ =\ -N (R+ {1\over2}\e\idone),
\eeq
where $R=\sum \pb_i\pi_i$ denotes the (diagonal)  matrix of $U(1)_R$ charges in the Chan-Paton space.
 Thus we can formally associate the normal ordering ambiguity with 
the grade of the $D$-brane, which too amounts to
an arbitrary overall shift of the $R$-charge.

Having defined a cohomological splitting on the space of matrices,
 we can now write a preferred basis of the physical operators in closed form:
 \beq
\ll{eq:goodbas}
\Psi_{i_1..i_k} :=\ \Pi_P\cdot \pi_{i_1}....\pi_{i_k}\ \in \cH_P\ , \ \ \ k\leq N.
\eeq
By construction they are BRST closed but not exact: $Q\cdot \Psi_{i_1..i_k}=0=\Udel\cdot \Psi_{i_1..i_k}$ (\ie, satisfy a Siegel-type gauge).
This basis is analogous to a polynomial basis $\{y_a(x)\}$ of Jac$(W)$ with $U(y)=0$, in the bulk theory.

To conclude this section, we hasten to clarify a potential confusion: 
for the operators (\ref{eq:goodbas}), where are the labels for the boundaries?  The point is
that LG models describe orbifolds of geometrical theories, with self-intersecting branes. \
As we will review later, the boundary labels of $\Psi_{i_1..i_k}^{(A,B)}$ appear only after un-orbifolding, 
which leaves the form of the operators invariant. 

\subsection{Joint Bulk-Boundary Deformation Problem}

In the following, we shall be interested in bulk deformations
of the open string TFT defined by the matrix factorization \eqz{Qsquare};
see~\cite{Herbst:2004zm,Hori:2004ja,Brunner:2004mt,Govindarajan:2005im,Knapp:2006rd,Govindarajan:2006uy,Knapp:2008uw,Baumgartl:2010ad,Carqueville:2009ay,Carqueville:2011aa,Carqueville:2011px} 
for some previous works on deformations of matrix factorizations.
We will denote the closed string moduli by $t_a$ as before and
possible open string moduli of a brane by
$u_\al$ (not to be confused with the spectral parameter $u$).
 With ``moduli'' we refer to operators with $R$-charge $q=2$ in the
bulk or $q=1$ on the boundary, so that $t_a$ and $u_\al$ are
dimensionless and can appear in correlation functions in a
non-polynomial way. On the other hand, we will denote relevant,
``tachyonic'' deformations coupling to boundary changing operators $\Psi_\al$ by $s_\al$; 
being dimensionful, they can appear in the effective potential
only in a polynomial way.

Strictly speaking, because in general an effective potential
$\cW_{eff}(t,u,s)$ will be generated, some or all of these deformations won't be true moduli but will be obstructed 
(possibly at higher order).  The true moduli space consists of the sub-locus of
the joint $(t,u,s)$ deformation space that preserves the factorization, \ie, $Q(t,u,s)^2=W(t)$. 
It can be shown \cite{Knapp:2006rd,Carqueville:2011aa} 
that this supersymmetry preserving locus coincides with the
critical locus, $\del_{u,s}\cW_{eff}(t,u,s)=0$, of the effective
superpotential.

In the following, we will consider only 
unobstructed, supersymmetry preserving factorization loci, and concentrate on bulk-deformed boundary
operators, $\Psi=\Psi(t)$; we will consider possible boundary moduli $u_\al$ as frozen.  More specifically, we will work
to all orders in the bulk perturbation but only infinitesimally in
the boundary changing open string sector:\footnote{For this to make sense if $\Psi$
is a boundary changing operator (which we assume), one should
view $Q$ as a block matrix containing $Q_A$ and $Q_B$ in the diagonal,
perturbed by the off-diagonal element $s_a\, \Psi_a^{(A,B)}$.}
\beq
\ll{eq:firstorder}
Q(t,s)\ =\ Q(t) + s_\al\,\Psi_\al(t)\ .
\eeq
Note that factorization is preserved to lowest order as long as $[Q(t),\Psi_\al(t)]=0$. 
An obstruction, and thus a superpotential, can arise only if several boundary changing operators are switched on simultaneously. Thus, for the purpose of determining ``flat'' representatives $\Psi=\Psi(t)$, we can restrict our considerations to linear order in $s$ and consider the theory as un-obstructed at this order (and thus on-shell, making computations well-defined in CFT).

We have seen in the previous section how to systematically construct  canonical representatives of the boundary cohomology, and
even obtain explicit expressions \eqz{goodbas} for them for the class of factorizations we consider.
However, in order to obtain actual correlation functions, there is more
left to do, namely the $t$-dependent renormalization $g(t)$ of the operators $\Psi(t)$ still
needs to be determined. This is in fact the main part of the work, which it is usually neglected in the discussion of
matrix factorizations.

Before we will discuss this in the next section, note that the operators that couple to unobstructed,
supersymmetry-preserving deformations parametrized by $t_a$ are not just given by
the flat bulk fields defined by $\phi_a(x,t)=\del_aW(x,t)$.  Indeed,
the reason \cite{Warner:1995ay} for introducing boundary degrees of
freedom in the first place had been to cancel the $B$-type supersymmetry
variation of the bulk superpotential $W$ that arises on world-sheets
with boundaries. This implies that the factorization-preserving
deformations of $W$ that we consider, must necessarily be accompanied by
simultaneous deformations of $Q$ by certain boundary operators,
$\gamma_a\in \cH_{U}$. Differently said, we deal here with a joint open/closed deformation problem 
where the deformations are locked to a common un-obstructed deformation locus via the 
factorization condition \eqz{Qsquare}.
By definition, these induced boundary counter terms are not cohomology elements but rather obey:
\beq
\ll{eq:gammavariation}
\big[\,Q(t),\gamma_a(t)\,\big]\ =\ \phi_a(t)\vert_{\del D}\idone\ .
\eeq
That the bulk modulus $\phi$ must be BRST exact at the boundary for the deformation to be un-obstructed,  also 
follows from abstract reasoning (in that degree-two deformations in Ext$^2(*)$ determine obstructions \cite{Fukaya:2001uc} and so 
must be trivial by assumption).  
From \eqz{gammavariation} and factorization it follows that
\beq
\ll{eq:gammadef}
\gamma_a(t)\ =\ \del_{t_a}Q(t)\,,
\eeq
up to BRST closed pieces. The point is that only the combined perturbation:
\beq
\ll{eq:combined}
\sum t_a\int \tilde \phi_a\equiv\ \, \sum 
t_a\left(\int_D\phi_a\lab2\idone-\int_{\del D}\gamma_a\lab1\right)\, ,
\eeq
is preserved by the total, $B$-type BRST operator at the boundary,
\beq
\ll{eq:Qtotal}
\cQ_{tot}\ =\ \cQ_W\big\vert_{\del D}\idone
+\big[\,Q,*\,\big]\, ,
\eeq
while the individual terms are not.  This follows from \eqz{gammavariation}
and the descent equation, $[\cQ,\phi\lab2] = \del_\bot\phi\lab1$,
where $\phi \lab1$ is a one-form along the boundary of the disk.
Physically, \eqz{combined} implements the subtraction of the ``Warner''
contact term  $\int_{\del D}\phi\lab1$ of $\phi$ with the boundary, and mathematically 
represents the natural invariant pairing on the relative (co-)homology
of the disk. 

The coupled bulk-boundary perturbation \eqz{combined} is thus a quite peculiar observable: 
despite neither of its individual building blocks belongs to the physical cohomology of the 
boundary theory, it deforms correlation functions.  
Whenever discussing perturbed correlation functions such as the one written in \eqz{opencorr}, one
should implicitly use such combined, ``relative'' perturbations to cancel the bulk-boundary contact terms.

A key role is again played by contact terms, in particular the contact terms
that arise when the insertion $\int_{\del D}\ga\lab1$ hits other operators
at the boundary:
\beq
[\cQ,\int_{\del D}\ga\lab1]\Psi\ =\ (\int_{\del D}\phi\lab1)\Psi + [\ga,\Psi]\ .
\eeq
This follows from the descent equation $[\cQ,\ga\lab1] = \phi\lab1+\del_\parallel\ga$.
The second term on the RHS must then be cancelled by the $t$-variation of $\Psi(t)$. 
Indeed, inverting the BRST operator in $\del_t[Q,\Psi]=0$ we can write $\del_t\Psi=-\Udel([\ga,\Psi])$,
whose $Q$-variation reproduces this term.
Note that $[\ga,\Psi]\in\cH_E$ and thus the inversion is well-defined.

More succinctly, defining $\Psi_g(t)\equiv g(t)\Psi(t)$ one might be tempted to write the following differential equation:
\bea
\ll{eq:boundaryDEQ}
\nabla_t^U\Psi_g(t) & =& {g'(t)\over g(t)}\Psi_g(t) ,  \ \ {\rm where}\\
\nabla_t^U &\equiv& {\del\over \del t}+ \Udel(\left[\ga\, ,*\right])\,,  \ \  \nabla_t^U: \cH_P\to \cH_P\,,\nn
\eea
and wonder whether this is already the differential equation we are after. 
 Actually, it is not,  which is no surprise given its tautological derivation. 
 First, the inversion of $Q$ is determined only up to BRST closed pieces in  $\cH_P\oplus\cH_E$, and
 thus $g'(t)$ is not really determined.
 Moreover, by recalling how $\Psi(t)$ is defined in \eqz{goodbas} in terms of $\Udel$, and by explicitly following
 the action of $\del_t$ through the latter's definition in \eqz{UfullI}, it turns out that
 \eqz{boundaryDEQ} is in fact an identity.  

This means that the contact term that arises if $\int\ga$ hits $\Psi$ is
 already taken care of by the action of $\del_t$, and so no constraint on $g(t)$ can be obtained just from  \eqz{boundaryDEQ}  alone.
 Rather, we expect that $g(t)$ is determined by the interplay between the integrated insertions \eqz{combined}
 and this contact term. See Fig.~\ref{fig:contactterms} for a pictorial summary of the contact terms.

\begin{figure}[t]  
\begin{center}
\includegraphics[width=15cm]{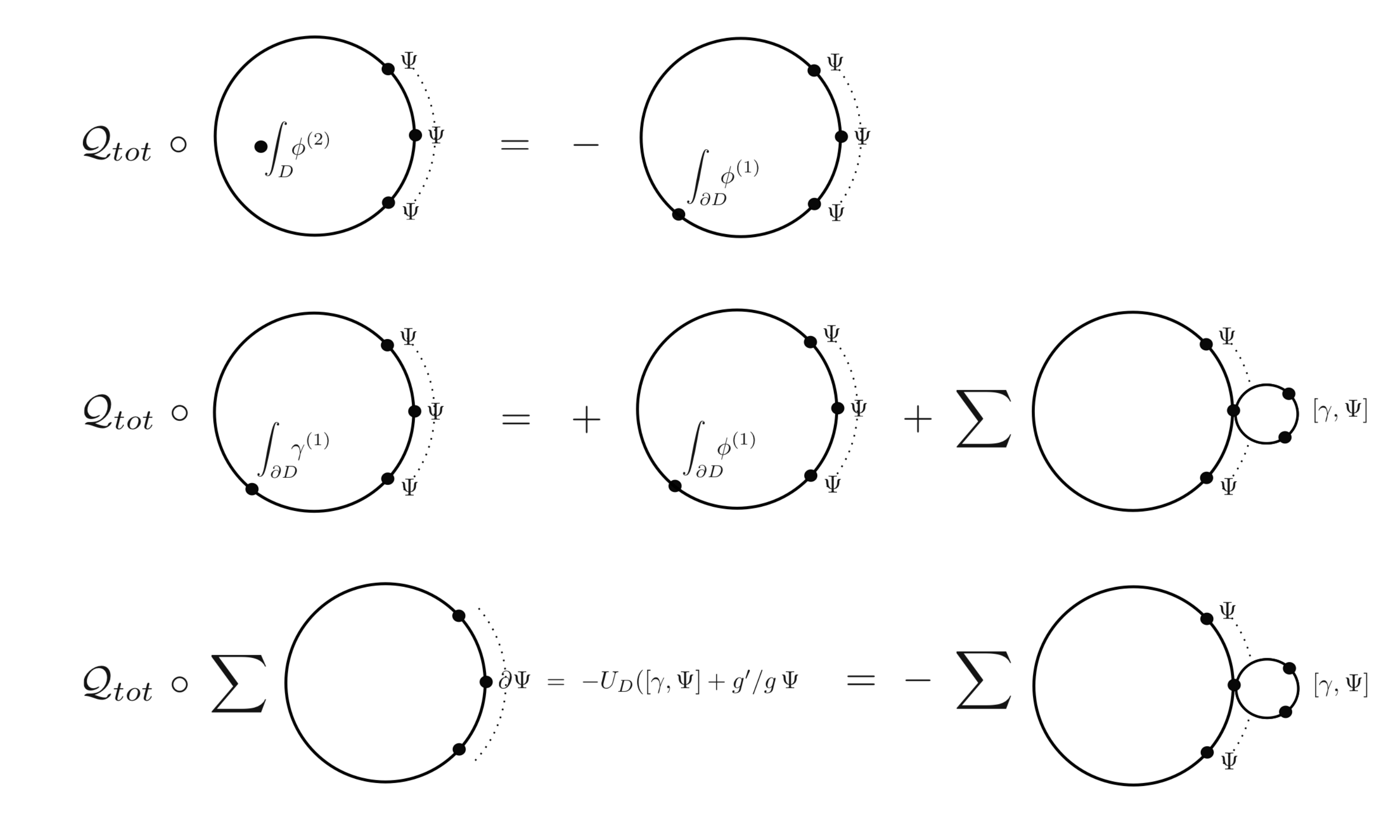}
\caption{Interplay between the bulk-boundary and boundary-boundary contact terms associated with
the combo-insertion \eqz{combined}. The total BRST variation cancels.
}
\ll{fig:contactterms}
\end{center}
\end{figure}

\subsection{Flatness and boundary pairings}

We now turn to a concrete realization of the desired flatness equations in LG language. For this we need to
consider correlation functions. The most important quantity is a generalization of the Grothendieck multi-residue pairing to
matrix factorizations, which was found by Kapustin and Li \cite{Kapustin:2002bi,Kapustin:2003ga,Segal:2009eg,Dyckerhoff:2010cm,Shklyarov:2013}.
It defines an inner product at the boundary as follows:
\beq
\ll{eq:KLori}
\langle \Psi_\al,\,\Psi_\be\rangle_{\del D}\  \ \equiv\ \eta_{\al\be}\ =\ K\lab0(1,\str[dQ^{\wedge n} \sdot\Psi_\al\sdot\Psi_\be])\ ,
\eeq
which is known to be non-degenerate.
This pairing is well-defined for cohomology elements, and maps $\cH_P\otimes\cH_P\to\bbC$.  This means that it
satisfies trace property and cyclicity of $n$-point functions only on-shell.
Off-shell where $\Psi_*\in\cH_U$, cyclicity is violated so this pairing does not define a Calabi-Yau structure proper \cite{Costello:2004}, 
without modifications. This problem has been addressed in refs.~\cite{Carqueville:2009ay,Shklyarov:2016}, 
where correction terms were determined that turn \eqz{KLori} into a good off-shell pairing.

In the following, we will not be concerned about off-shell properties of the pairing.
For now, focusing only on on-shell properties, we like to rewrite the Kapustin-Li supertrace-residue pairing in a symmetric form as follows:
\bea
\ll{eq:KL}
K\lab0_{KL}(\Psi_\al,\Psi_\be)\!\! \!&\equiv &\!\! \!  K\lab0_{KL}(1; \Psi_\al,\Psi_\be)\ ,\qquad{\rm with}\ \\  
K\lab0_{KL}(\xi; \Psi_\al,\Psi_\be) 
\!\! \!&= &\!\! \!
 K\lab0(\xi,\sum_k\str [dQ^{\wedge k}\sdot \Psi_\al \sdot dQ^{\wedge (n-k)}\sdot\Psi_\be])\\
\!\!\!\!&\equiv&\!\!\! {(-1)^n\over (n+1)!}\sum_{k=0}^n(-1)^{k|\Psi_\al|} \sum_{i_*=1}^n  \e_{i_1...i_n}\oint\xi\,\str
\left[{d_{i_1}Q\over d_{i_1}W}\dots {d_{i_k}Q\over d_{i_k}W}\,\Psi_\al
{d_{i_{k+1}}Q\over d_{i_{k+1}}W}\dots {d_{i_n}Q\over d_{i_n}W}\Psi_\be
\right]\!,\nn
\eea
where $|\Psi|$ denotes the $\bbZ_2$-grade of $\Psi$. This form obtains naturally
when one performs a derivation from the path integral \cite{Herbst:2004ax}, analogous to what we outlined in Section (2.2) for the bulk theory.

In view of our previous discussion, an obvious question is how to extend this to higher pairings, for example, by
introducing the spectral parameter $u$.
Such an extension has been constructed by Shklyarov \cite{Shklyarov:2013}. 

Note however that the spectral parameter $u$ has degree (or charge) 2, which is matched to the superpotential $W$ and so matches 
insertions of $d/dW$ in the higher residue pairings. On the other hand, at the boundary the relevant 
cohomology is determined by $Q$, which has degree 1,
so inversions of $Q$ (resp.\ contact terms) should formally be counted in terms of a degree 1, and not a degree 2 variable. 
This is closely tied to the fact that bulk deformations involve two fermionic integrations in $\int_D\phi\lab2$, while boundary deformations
involve only one in $\int_{\del D}\Psi\lab1$. That is, the spectral parameter $u$ seems to be a genuine bulk quantity that does not capture
all aspects of the boundary theory. It appears that for our purposes we need a different extension of \eqz{KL}, not in the direction $u$, 
but rather in a different, morally speaking anti-commuting direction.  

Before we will discuss this, let us mention an instance where an extension in the $u$-direction
appears to be useful for our purposes: namely we can consider an intermediate "bulk-boundary" pairing of the form
\beq
\ll{eq:Kbb}
K\lab\ell_{bb}(\varphi_a,\Psi_\be)\ :=\   
 K\lab\ell(\varphi_a,\str[dQ^{\wedge n}\sdot\Psi_\be]):\, \cH_{closed}\otimes \cH_{open}\ \to \bbC.
\eeq
That is, we treat the image of the open-closed map,
\beq
\ll{eq:OC}
OC[*]\ \equiv\ \str[dQ^{\wedge n} *]\!:\,\cH_{open}\to \cH_{closed}, \  \ \cH_P\to {\rm Jac}(W),
\eeq
like any other bulk operator. In analogy to \eqz{goodbas} we may propose as further condition for a ``good'' basis of operators
\bea
\ll{eq:goodbb}
K\lab0_{bb}(\varphi_a,\Psi_\be)\ &=& \eta_{a\be}\ =\  {\rm const}. \\
K\lab{\ell>0}_{bb}(\varphi_a,\Psi_\be)\ &=& 0\ .\nn
\eea
While we do not have a proof for this requirement, we will see later that these conditions 
make good sense at least in the context of an explicit example.

However note, importantly, that the bulk-boundary pairing \eqz{Kbb} affects only boundary preserving operators: $\Psi=\Psi\lab{A,A}$.
As mentioned in the introduction, this is  because the open-closed map vanishes identically on boundary changing operators, 
$\Psi\lab{A,B}$ with $A\not=B$.
Thus $K\lab\ell_{bb}$ is insensitive to the intrinsically new features of boundaries, 
namely  the ones that cannot be mapped to the bulk theory. 

Let us return to our task and try to find a pairing that is useful for our purpose, namely ultimately formulating flatness equations.
We will be guided by considering deformations of correlators in LG language, in analogy to what we have reviewed for the bulk theory.
That is, requiring constancy of the inner product  \eqz{KLori} yields
\bea
\ll{eq:boundconst}
0 &=& {\del\over \del t} K\lab0_{KL}(\Psi_\al,\Psi_\be)\\
&=& K\lab0_{KL}({\del\over\del t} \Psi_\al,\Psi_\be)+ K\lab0_{KL}( \Psi_\al,{\del\over\del t}\Psi_\be) 
- K\lab0_{KL}(\sum_i {d_i\phi\over d_iW}; \Psi_\al,\Psi_\be)
\nn\\
&+&  K\lab0\Big(1,\sum_k\str\big[ \big({\del\over \del t}dQ^{\wedge k}\big)\sdot \Psi_\al \sdot dQ^{\wedge (n-k)}\sdot\Psi_\be\big]\Big)
+  K\lab0\Big(1,\sum_k\str\big[ dQ^{\wedge k}\sdot \Psi_\al \sdot \big( {\del\over \del t}dQ^{\wedge (n-k)}\big)\sdot\Psi_\be\big]\Big).\nn
\eea
The new ingredient is the action of $\del_t$ on the $dQ$'s. This is in line of what we discussed in the previous section, namely
that we deal here with a coupled bulk-boundary deformation problem. Eq.\ \eqz{boundconst} explicitly
manifests in LG language the heuristic correspondence:
\beq
\big(\int_D\phi\lab2\idone-\int_{\del D}\gamma\lab1\big) \ \ \ \longleftrightarrow\ \ \ \big({d\phi\over dW}- {d\gamma\over dQ}\big).
\eeq
Here ``$d\ga/dQ'$'' has the meaning to drop a $dQ$ and replace it with its $t$-derivative, in all the proper locations.\footnote{Recall that the $dQ$'s and thus $d\ga$'s can be different for the two boundary segments, namely if these are associated with different matrix factorizations,
$\cM(A)$ and $\cM(B)$. Relatedly, implicit in the notation $\int_{\del D}\gamma\lab1$  is that  there can be different boundary segments of the disk $D$, and the insertions must be done
according to the respective boundary conditions.}

We presented our problem in a way that suggests a generalization of the flatness equations as follows. 
We see from the structure of \eqz{boundconst} that the kind of higher boundary pairing we are after, 
should involve $d/dQ$ (of degree $-1$) instead of $d/dW$ (degree $-2$). Thus we are lead to propose
 as first higher supertrace-residue pairing:
\bea
\ll{eq:boundpair}
K\lab1_{KL}(\Psi_\al,\Psi_\be)&:=&
{(-1)^{n+1}\over (n+1)!}\sum_{k=1}^{n}(-1)^{k(|\Psi_\al|+1)} \sum_{i_*=1}^n  \e_{i_1...i_n} \times \\
&&{2}\oint\str
\Bigg[{\left({d_{i_1}Q\over d_{i_1}W}\dots {d_{i_{k-1}}Q\over d_{i_{k-1}}W}{d_k\Psi_\al\over d_kW}
{d_{i_{k+1}}Q\over d_{i_{k+1}}W}\dots {d_{i_n}Q\over d_{i_n}W}\, \,\Psi_\be\right) }  
\nn\\
&&\qquad\qquad-
\left({d_{i_1}Q\over d_{i_1}W}\dots {d_{i_k}Q\over d_{i_k}W}\,\Psi_\al
{d_{i_{k+1}}Q\over d_{i_{k+1}}W}\dots {d_{i_{n-1}}Q\over d_{i_{n-1}}W}{d_n\Psi_\be\over d_nW}\right)
\Bigg],\nn
\eea
which satisfies $K\lab1_{KL}(\Psi_\al,\Psi_\be)=(-1)^{|\Psi_\al||\Psi_\be|+1} K\lab1_{KL}(\Psi_\be,\Psi_\al)$.
In terms of this we can rewrite equation \eqz{boundconst}  as
\bea
\ll{eq:boundconst1}
0 &=& {\del\over \del t} K\lab0_{KL}(\Psi_\al,\Psi_\be)\nn\\
&=&- K\lab0_{KL}(\sum_i {d_i\phi\over d_iW}; \Psi_\al,\Psi_\be)\\
&&+\ K\lab0_{KL}({\del\over\del t} \Psi_\al,\Psi_\be)+ K\lab0_{KL}( \Psi_\al,{\del\over\del t}\Psi_\be)\nn\\
&&+\ K\lab1_{KL}(\ga\sdot\Psi_\al,\Psi_\be)+ K\lab1_{KL}( \Psi_\al,\ga\sdot\Psi_\be)\ .\nn
\eea
This is supposed to be the boundary analog of eq.\ \eqz{constmetric}  in the bulk theory.

In fact, this equation represents an identity:
the cohomology elements as defined in \eqz{goodbas} are already normalized such that all inner products are constant,
if the overall normalization is chosen such the one-point function of the top element is constant: 
$K\lab0_{KL}(\Psi_{12..N},\idone)=K\lab0(H,1)=1$. Thus,  eq.\ \eqz{boundconst1} by itself does not have too a great significance.

Indeed, requiring that the topological boundary two-point function \eqz{KLori}
\beq
\ll{eq:binner}
\langle\,\Psi^{(A,B)}_\al\,\Psi^{(B,A)}_\be\,\rangle_{\del D}\ =\ \eta_{\al\be}\ ,
\eeq
be constant, is of little help for fixing the operators, 
because it is invariant under the relative
rescaling $\Psi^{(A,B)}\to g(t)\Psi^{(A,B)}$, $\Psi^{(B,A)}\to g(t)^{-1}\Phi^{(B,A)}$. 
Thus, we cannot determine the independent renormalization factors from
\eqz{boundconst1}. However we need to know them because, for example, the three-point
function $\langle \Psi^{(A,B)}\Psi^{(B,C)}\Psi^{(C,A)}\rangle$ will be proportional to $g(t)^3$. 
This is why we need to find differential equations that determine the relative
moduli-dependent renormalization factors for all fields individually, and not just of their products. 
It also relates back to what we said in the Introduction: as long as we consider only
closed cycles of operators, $\Psi^{(A,B)}\Psi^{(B,C)}\dots \Psi^{(*,A)}\in HH_*(Cat(M\!F,W))$, we  
cancel out important information, namely the one that intrinsically goes beyond the bulk theory.

Thus all boils down to one basic and crucial problem, 
namely to as to how to split equation \eqz{boundconst1} into two separate, stronger ones.
Without any deeper insights, this is an ambiguous problem, namely what forbids us to add and subtract terms
to the individual pieces such that they cancel out in the sum? Alarmingly, in the end, 
practically all correlations functions that we want to compute will depend on this split!

The only pragmatic way we see, is to be guided by analogy of the bulk (cf., \eqz{fullflatness}) and by the structure of \eqz{boundconst1},
and define ``relative'' connections by
 \bea
\ll{eq:finalD}
 K\lab0_{KL}(\nabla_t\Psi_\al,\Psi_\be)\! \!\!&:=& \! \!\! K\lab0_{KL}(\del_t\Psi_\al,\Psi_\be)+ K\lab1_{KL}(\Psi_\al,\ga\sdot\Psi_\be)
 -{1\over2} K\lab0_{KL}(\sum_i {d_i\phi\over d_iW}; \Psi_\al,\Psi_\be)
 \\
  K\lab0_{KL}(\Psi_\al,\nabla_t\Psi_\be)\!\!\!&:=&\!\!\!  K\lab0_{KL}(\Psi_\al,\del_t\Psi_\be)+ K\lab1_{KL}(\ga\sdot\Psi_\al,\Psi_\be)
 -{1\over2} K\lab0_{KL}(\sum_i {d_i\phi\over d_iW}; \Psi_\al,\Psi_\be)
  .\nn
  \eea
It is convenient to rewrite the differential equations into a common mode and a relative mode part as follows:
 \bea
\ll{eq:finalDEQ}
K\lab0_{KL}(\nabla_t\Psi_\al,\Psi_\be) + K\lab0_{KL}(\Psi_\al,\nabla_t\Psi_\be) &=& 0\,\\
 \ll{eq:finalDEQ1}
 K\lab0_{KL}(\nabla_t\Psi_\al,\Psi_\be) - K\lab0_{KL}(\Psi_\al,\nabla_t\Psi_\be) &=& 0\,.
 \eea
This is what we propose, without proof, for an explicit realization of the flatness equations \eqz{flatparing} that we advertized in the Introduction.
 As said, the first equation can be satisfied by a judicious common mode normalization, which we assume
  (ratios of correlators will be invariant under changes of the overall normalization, anyway).
 The more non-trivial, new information is in the second equation \eqz{finalDEQ1} which samples the relative normalization of the operators.
  
 A few remarks are in order.  
 
First, note that the covariant derivatives \eqz{finalD}
are manifestly written in terms of integrated insertions only. 
As mentioned in the previous section, the contact terms between $\ga$ and $\Psi$ are already 
taken care of by the $t$ derivatives acting on the $\Psi$'s. One may wonder whether
there could be additional, contact terms directly between $\phi$ and the $\Psi$'s as well. In fact, it is known that the possible stable
degenerations of the punctured disk do not include such factorizations, rather degenerations involving bubbling off disks appear only when
boundary operators hit each other; see again {Figure}~\ref{fig:contactterms}. So $\phi$ can enter only as integrated operator, 
which means it should enter symmetrically with respect to the $\Psi$'s, precisely as it does
in \eqz{finalD}.

To see the structure of the differential equations more clearly, it is helpful to disentangle
 the scalar-valued renormalization of the operators, $g(t)$, from the boundary contact term. 
 Using \eqz{boundaryDEQ} we can rewrite
 \bea
 \ll{eq:finalD1}
 && \!\!\!\!\!\!\!\!\!\!\!\!\!K\lab0_{KL}(\nabla_t\Psi_\al,\Psi_\be)  
 \\
 &&=K\lab0_{KL}({g'(t)\over g(t)}\Psi_\al,\Psi_\be)- K\lab0_{KL}(\Udel([\gamma,\Psi_\al],\Psi_\be)+K\lab1_{KL}(\Psi_\al,\ga\sdot\Psi_\be)
  -{1\over2} K\lab0_{KL}(\sum_i {d_i\phi\over d_iW}; \Psi_\al,\Psi_\be),\nn
  \eea
  where we have implicitly rescaled $\Psi_\al$ by $g(t)$ and $\Psi_\be$ by $1/g(t)$.
This exhibits the interplay between the contact term of $\gamma$ with $\Psi_\al$,  and the integrated insertions.
 In a sense, the renormalization factor $g(t)$ is determined by a mismatch between these terms.

 Second, it may appear counter-intuitive that the middle 
 terms on the RHS in \eqz{finalD} look switched. Actually being not contact terms but
 integrated insertions, there is no reason why $\ga$ should stick locally to the operators on which we take derivatives.  From their origin,
 they sample the $dQ's$ rather then the $\Psi$'s. One can check that
precisely the combinations given in \eqz{finalD} have good covariance  properties under transformations  
 $Q\to V(t)^{-1}QV(t)$. This will become evident in the example that we will discuss below.
 Also, note that the ordering of $\ga$ with respect to the $\Psi$'s does not matter, up to signs.
 
Finally, what about higher pairings $ K\lab\ell_{KL}$ for $\ell>1$? Higher versions with more derivatives acting on the $\Psi$'s
can be constructed in analogy to \eqz{boundpair}.
Due to the anti-commuting nature of this kind of pairings and the limited number of $dQ$'s,
it is clear that there can be just a finite number of them. Such higher versions would play a role in a more thorough treatment.
However, at this point we are not sufficiently certain about the mathematical logic
of such higher pairings, so we prefer to leave this issue to later work.
For now, we content ourselves to test the proposed equation \eqz{finalDEQ1} for the simplest possible case, 
namely for the cubic elliptic curve.

\subsection{Example: the elliptic curve revisited}

\subsubsection{$B$-Model computations}

We now re-visit open string mirror symmetry for the cubic elliptic curve. In the physics literature
this has been discussed in refs.~\cite{Hori:2004zd,Brunner:2004mt,Govindarajan:2005im,Govindarajan:2006uy,Knapp:2007kq}.
We consider the canonical matrix factorization \eqz{Qcan}
\bea
\ll{eq:Qdef}
Q(x,t)&=&  1/3\pi_i x_i + \pb_i d_iW(x,t)\ ,  \  \  Q(x,t)\cdot Q(x,t)=W(x,t),\\
\ll{eq:Wdef}
W(x,t)&=& \ft(t)\left[{1\over 3}\sum{x_i}^3- \alpha(t) x_1x_2x_3\right] ,
\eea
with   $ \{\pi_i,\bar\pi_j\}=\delta_{ij}$, $i,j=1,2,3$.
 It corresponds to a special, irreducible point of a continuous family of otherwise reducible
matrix factorizations; for details see ref.~\cite{Govindarajan:2005im}. 
This means that the open string moduli $u_\al$ (locations of $D0$ branes)
are frozen, and can be put to zero
in a suitable coordinate system. We will give some more geometrical information later.

Recall that a canonical basis of the non-trivial cohomology elements is given by \eqz{goodbas}, which by construction
are in a generalized Siegel gauge, \ie, obey $\Udel\cdot \Psi_{i_1..i_k}=0$. Let us put the relative 
renormalization factors that we need to determine, as follows:
\bea
\ll{eq:Pdef}
\Psi(t) &=& g_0(t)^{-1}\,\idone\nn\\
\Psi_i(t)&=& g_1(t)\, \Pi_P\cdot \pi_i\\
\Psi_{ij}(t)&=& g_1(t)^{-1}\, \Pi_P\cdot\pi_i\pi_j,\nn \\
\Psi_{ijk}(t)&=& g_0(t)\, \Pi_P\cdot\pi_i\pi_j\p_k.\nn
\eea
The top element has charge $1$ and represents a marginal operator that couples to the $D0$ brane modulus $u$
(which we suppress).
We will need only a few of these operators explicitly:
\bea
\ll{eq:psis12}
g_1(t)^{-1}\Psi_1(t)&=&\pi_1-3 \ft x_1 \bar\pi_1 + \frac{3}{2}\ft \al\Big[ (x_3 \bar\pi_2+ x_2 \bar\pi_3 )\Big],
\\
g_1(t)\Psi_{23}(t)&=&
\pi_2\pi_3 
+\frac{9}{4}\al \ft^2 
\Big[  \left( 2 x_2{}^2 +\ x_1 x_3 \al\right)\bar\pi_1\bar\pi_2
-\left( 2x_3{}^2+x_1 x_2 \al\right)\bar\pi_1\bar\pi_3 \Big]
   \\
&&   +9 \ft^2\big( x_2 x_3 -\frac{1}{4} x_1{}^2 \al^2 \big)\bar\pi_2\bar\pi_3 
-3 x_2 \ft \bar\pi_2\pi_3+3 x_3 \ft \bar\pi_3\pi_2
\nn   \\
   &&+\frac{3}{2} \al \ft \Big[ x_3 \bar\pi_1\pi_3- x_2\bar\pi_1\pi_2
  + x_1 \bar\pi_3\pi_3 - x_1 \bar\pi_2\pi_2
  \Big], \nn
 \eea
   and
\bea
\ll{eq:psis123}
g_0(t)^{-1}\Psi_{123}(t)&=&   
\pi_1\pi_2\pi_3 -\frac{27}{8} H  \bar\pi_1\bar\pi_2\bar\pi_3
   \\
&& \!\!\!\!\!\!\! \!\!\!\!\!\!\! \!\!\!\!\!\!\! \!\!\!\!\!\!\!
+\frac{9}{4} \al \ft^2 \Big[\left(x_1 x_3 \al+2 x_2{}^2\right) \bar\pi_1\bar\pi_2\pi_1
+ \left(x_2 x_3 \al+2 x_1{}^2\right) \bar\pi_1\bar\pi_2\pi_2
   -\left(x_1 x_2 \al+2 x_3{}^2\right) \bar\pi_1\bar\pi_3\pi_1
\nn\\
&&   -\left(x_2 x_3 \al+2 x_1{}^2\right) \bar\pi_1\bar\pi_3\pi_3
 +\left(x_1 x_2 \al+2 x_3{}^2\right) \bar\pi_2\bar\pi_3\pi_2
   + \left(x_1 x_3 \al+2 x_2{}^2\right) \bar\pi_2\bar\pi_3\pi_3
   \Big]
\nn   \\
&& \!\!\!\!\!\!\! \!\!\!\!\!\!\! \!\!\!\!\!\!\! \!\!\!\!\!\!\!
+\frac{9}{4} \ft^2\Big[ 
\left(4 x_1 x_2-x_3{}^2 \al^2\right)   \bar\pi_1\bar\pi_2\pi_3
+\left(4 x_2 x_3-x_1{}^2 \al^2\right) \bar\pi_2\bar\pi_3\pi_1
 -\left(4 x_1 x_3-x_2{}^2 \al^2\right) \bar\pi_1\bar\pi_3\pi_2
     \Big]
\nn   \\
   && \!\!\!\!\!\!\! \!\!\!\!\!\!\! \!\!\!\!\!\!\! \!\!\!\!\!\!\!
 +\frac{3}{2}\al \ft\Big[ 
x_2  \bar\pi_1\pi_1\pi_2- x_3  \bar\pi_1\pi_1\pi_3  + 
x_1 \bar\pi_2\pi_1\pi_2  +\ x_3  \bar\pi_2\pi_2\pi_3
- x_1 \bar\pi_3\pi_1\pi_3+ x_2 \bar\pi_3\pi_2\pi_3
\Big]
\nn   \\
   &&-3 \ft\Big[
   x_1  \bar\pi_1\pi_2\pi_3+ x_2  \bar\pi_2\pi_1\pi_3- x_3 \bar\pi_3\pi_1\pi_2
   \Big]. \nn
\eea
We also do some modifications: analogous to the rescaling by $\ft(t)$ of the superpotential $W(x,t)$,
we have an corresponding degree of freedom at the boundary, namely a rescaling of the boundary fermions as follows:
\beq
\ll{eq:boudaryresc}
\pi_i \ \rightarrow \rt(t)\pi_i\,,\qquad \bar\pi_i \ \rightarrow \rt(t)^{-1}\pb_i\,.
\eeq
This in particular affects the boundary counter term as follows:
\beq
\ll{eq:gammar}
\ga(t)\equiv{\del\over \del t}Q(t)\ =\ \Udel(\phi(t) \idone)-3 {\rho'(t)\over\rho(t)} \big[Q, R]\,,
\eeq
where $R$ is the matrix of $R$-charges and  $\phi$ is as given in \eqz{phidef}. 
Moreover, in analogy to the bulk theory, where $\phi$ had to be shifted by an exact piece (consistent with the condition
of good basis \eqz{goodbas}), we allow for a shift of $\Psi_{123}$ by an exact piece. It turns out that given the various
possible $Q$-exact parameters, their image in the various residues is only one dimensional. So we write this extra exact piece
conveniently in terms of just one  parameter, $\lambda(t)$:
\bea
\Psi_{123}(t)& \to& \Psi_{123}(t)-  \lambda(t)\,\big[Q,\Lambda\big],\\
\Lambda &=&  27^2 g_0(t)\rt(t)^{-2} x_2\, \bar\pi_1\bar\pi_2\bar\pi_3\pi_2.
\eea

We now assemble correlation functions from the Kapustin-Li and related pairings.  First we rescale them like in the bulk, \ie,
$\tilde K^{(*)}_{*}(..,...) :\equiv\ft^{-3} K^{(*)}_{*}(..,...)$, up to a constant factor. In particular,
\beq  
\ll{eq:normalizb}
\langle\, \Psi_\al,\Psi_\be \rangle_{D}\  \equiv\ \langle\langle\Psi_\al,\Psi_\be \rangle\rangle\ =\ \tilde K^{(0)}_{KL}(\Psi_\al,\Psi_\be) .
\eeq
It is easy to check the constancy of the boundary inner products:
\beq
\ll{eq:btwopoint}
\langle\, \Psi(t)\Psi_{ijk}(t) \rangle_{D} \ =\ \langle\, \Psi_{i}(t)\Psi_{jk} (t)\rangle_{D}\ =\ \e_{ijk}\ .
\eeq
Moreover we have for all operators $\tilde K^{(1)}_{KL}(\Psi_{*},\Psi_{*})=0$.

We now compute the various pairings explicitly, and for this it suffices to consider the  operators in eqs.\ 
(\ref{eq:psis12}-\ref{eq:psis123}). Let us start with $\Psi_{123}$ and consider the integrated bulk insertion first:
\bea
\ll{eq:Kbb0}
 K\lab0_{KL}(\sum_i {d_i\phi\over d_iW};  \Psi_{123},\Psi)
&=& 3\frac{ \ft'(t)}{\ft(t)}+\frac{9 \al^2 \al'(t)}{4\Delta}- 3\lambda(t) 
\nn\\
&=&  6  \frac{ \eta'(t)}{\eta(t)}- 3\lambda(t) .
\eea
Something nice happens here, namely $\al(t)$ and $\ft(t)$ 
conspire such as to produce the Dedekind function ($q=e^{2\pi i t}$):
\bea
\eta(t)& \equiv & q^{1/24}\prod_{n>0}(1-q^n)(t) \ =\ {(-\sqrt 3\,\al'(t))^{1/4}\over \Delta^{1/8}}, \ \ \ {\rm therefore:} \\
{\ft'(t)\over \ft(t)}   &=&   2 {\eta'(t)\over \eta(t)}  -{3 \al(t)^2\al'(t)\over 4\Delta} \nn\ .
\eea
The nicety continues for all the other pairings:
\bea
\ll{eq:pairings0}
\tilde K^{(0)}_{KL}(\del_t\Psi_{123},\,\Psi)&=& \frac{g_0'(t)}{g_0(t)}+\frac{\rho'(t)}{\rho(t)}+2\frac{ \eta '(t)}{\eta (t)}-\lambda (t),  \\
\tilde K^{(0)}_{KL}(\Psi_{123}\,,\del_t\Psi)&=&  -\frac{g_0'(t)}{g_0(t)}, \nn \\
\tilde K^{(1)}_{KL}(\ga\sdot\Psi_{123},\,\Psi)&=&  2\frac{ \eta '(t)}{\eta (t)}-3 \lambda (t) ,\nn\\
\tilde K^{(1)}_{KL}(\Psi_{123},\ga\sdot\Psi)&=&2 \frac{ \eta '(t)}{\eta (t)}-\frac{\rho'(t)}{\rho(t)}+\lambda (t)  \nn
\eea
Thus in the covariant derivatives \eqz{finalD} the $\rho$-dependence cancels out:
\beq
\ll{eq:covs0}
\tilde  K\lab0_{KL}(\nabla_t\Psi_{123},\,\Psi) \ =\ \frac{g_0'(t)}{g_0(t)}+\frac{\eta '(t)}{\eta (t)} + \frac32 \lambda (t)
\ =\ 
-\tilde  K\lab0_{KL}(\Psi_{123},\,\nabla_t\Psi).
\eeq
Their sum cancels; thus the first differential equation \eqz{finalDEQ} is satisfied identically, as expected.
More interesting is the difference which depends on $\lambda (t)$:
\beq
\ll{eq:dif}
\tilde  K\lab0_{KL}(\nabla_t\Psi_{123},\,\Psi)-\tilde  K\lab0_{KL}(\Psi_{123},\,\nabla_t\Psi)\ =\ 
2\frac{g_0'(t)}{g_0(t)}+2\frac{\eta '(t)}{\eta (t)}+3 \lambda (t) 
\eeq
This is analogous to what happened in the bulk theory, where $\phi$ had to be shifted by an exact piece in order
to obtain a trivial relative normalization factor between $1$ and $\phi$. 
To fix $\lambda$, we invoke the proposed bulk-boundary pairing conditions  \eqz{goodbb}:
\bea
\ll{eq:btextra}
K^{(0)}_{bb}(1,\Psi_{123}) &=& g_0(t),
\\
\tilde K^{(1)}_{bb}(\phi,\Psi_{123})\ &=&g_0(t) \Big(2 \frac{ \eta '(t)}{\eta (t)}+3 \lambda (t)\Big). \nn
\eea
The second equation determines $\lambda (t)$ and so the differential equation finally turns into:
\beq
\ll{eq:g0zero}
g_0'(t)\ =\ 0,
\eeq
precisely as required. This then also fixes the first condition in \eqz{btextra}. We thus see some degree of
consistency of the procedure.

Now on to the more interesting sector, where we find:
\bea
\ll{eq:pairings1}
\tilde K^{(0)}_{KL}(\del_t\Psi_{1},\Psi_{23})&=&  \frac{g_1'(t)}{g_1(t)}+\frac{\rho'(t)}{3 \rho(t)}+\frac{2 \eta '(t)}{3 \eta (t)}  \\
\tilde K^{(0)}_{KL}(\Psi_{1},\del_t\Psi_{23})&=&  -\frac{g_1'(t)}{g_1(t)}+\frac{2 \rho'(t)}{3 \rho(t)}+\frac{4 \eta '(t)}{3 \eta (t)} \nn \\
\tilde K^{(1)}_{KL}(\ga\sdot\Psi_{1},\Psi_{23})&=&  \frac{2 \eta '(t)}{\eta (t)}-\frac{2 \rho'(t)}{3 \rho(t)}  \nn\\
\tilde K^{(1)}_{KL}(\Psi_{1},\ga\sdot\Psi_{23})&=&  \frac{2 \eta '(t)}{\eta (t)}-\frac{\rho'(t)}{3 \rho(t)} \nn
\eea
and so
\bea
\ll{eq:covs1}
\tilde  K\lab0_{KL}(\nabla_t\Psi_{1},\,\Psi_{23}) \ =\ \frac{g_1'(t)}{g_1(t)}-\frac{\eta '(t)}{3 \eta (t)}
\ =\ -\tilde  K\lab0_{KL}(\Psi_{1},\nabla_t\Psi_{23}).
\eea
Again, $\rho(t)$ cancels and the sum of both equations cancels. 
This then finally determines, up to a multiplicative constant:
\beq
\ll{eq:g1thirdt}
g_1(t)\ =\ \eta(t)^{1/3}\ .
\eeq

\subsubsection{$B$-Model correlators}

With the flattening renormalization factors at hand, we are ready to compute correlation functions. The simplest one is
\beq
\ll{eq:etacorr}
\langle\, \Psi_i\Psi_{j}\Psi_k\rangle_{D} \ \equiv\
\langle\langle\, \Psi_i,m_2(\Psi_{j} \otimes\Psi_k\rangle\rangle\  =\ \eta(t)\  \e_{ijk},
\eeq
where the lowest $A_\infty$ product is just matrix multiplication in the chiral ring:
\beq
\ll{eq:m2}
m_2(\Psi_j \otimes\Psi_k)\  \equiv\ \Psi_j \sdot\Psi_k\ =\  \eta(t)\, \Psi_{jk} + [Q,*]\ .
\eeq
Before we will discuss its significance below, let us first consider higher point correlators.

As pointed out in the introduction \eqz{mopencorr}, higher point correlators involve higher $A_\infty$ products.
These can be recursively assembled in terms of the boundary
propagator $\Udel$ defined in \eqz{UfullI} and lower products, forming nested trees such as
in {Figure}~\ref{fig:HMS}.   The functional complexity, on the other hand, is largely governed by the proper ``flat'' renormalization 
factors, which are usually neglected in this context.

The first non-trivial $A_\infty$ product is defined by
\beq
\ll{eq:mthree}
m_3(\Psi_\al\otimes\Psi_\be \otimes\Psi_\ga)\ =\ \Udel(\Psi_\al\sdot\Psi_\be )\sdot\Psi_\ga-
 (-1)^{|\Psi_\al|} \Psi_\al\sdot\Udel(\Psi_\be\sdot\Psi_\ga ),
\eeq
whose $Q$-variation measures the non-associativity of projected OPE's:
\beq
\ll{eq:Qm3}
[Q,m_3(\Psi_\al\otimes\Psi_\be \otimes\Psi_\ga)]\ = \Pi_P(m_2(\Psi_\al\otimes\Psi_\be))\sdot \Psi_\ga -
 \Psi_\al\sdot\Pi_P(m_2(\Psi_\be\otimes\Psi_\ga)).
\eeq
With this a particularly nice correlator can be computed explicitly, by inserting  
a ``weak bounding chain'' \cite{Fukaya:2001uc}  $\Psi_s\equiv -1/3 \sum s_i \Psi_i$ into \eqz{mthree}. This yields
\beq
\ll{eq:mthrees}
m_3(\Psi_s\otimes\Psi_s \otimes\Psi_s)\ =\  \eta(t)\,W(s,t)\,\idone\ ,
\eeq
where $W(x,t)$ is the LG superpotential defined in \eqz{Wdef}. 
The interpretation of this is that these are the two non-vanishing terms of the Maurer-Cartan equation \eqz{MCs},
where the zeroth product: $m_0=- \eta(t)\,W(s,t)\,\idone$ represents the curvature term of the deformed $A_\infty$ algebra.
Such a solution of the Maurer-Cartan equation with non-zero $m_0$ is called ``weakly obstructed".

Going back to physics, note that product $m_3$ in \eqz{mthrees} leads to the following correlator
\bea
\ll{eq:bfourpt}
\langle\langle\, \Psi_{123},m_3(\Psi_s\otimes\Psi_s \otimes\Psi_s)\rangle\rangle  =\ 
\langle \int\Psi_{123}\lab1 \Psi_s\Psi_s\Psi_s\rangle_{D}
&=&    {\del\over \del u}\cW_{eff}(s,t,u)|_{u=0}\
\\
&=&   \eta(t)\,W(s,t)\ ,\nn
\eea
where $u$ is the open string modulus that couples to the marginal boundary operator $\Psi_{123}$.
This describes an obstruction that appears if all three $s_i$ are switched on simultaneously.

\subsubsection{Instantons in the $A$-Model}

\begin{figure}[t]  
\begin{center}
\includegraphics[width=10cm]{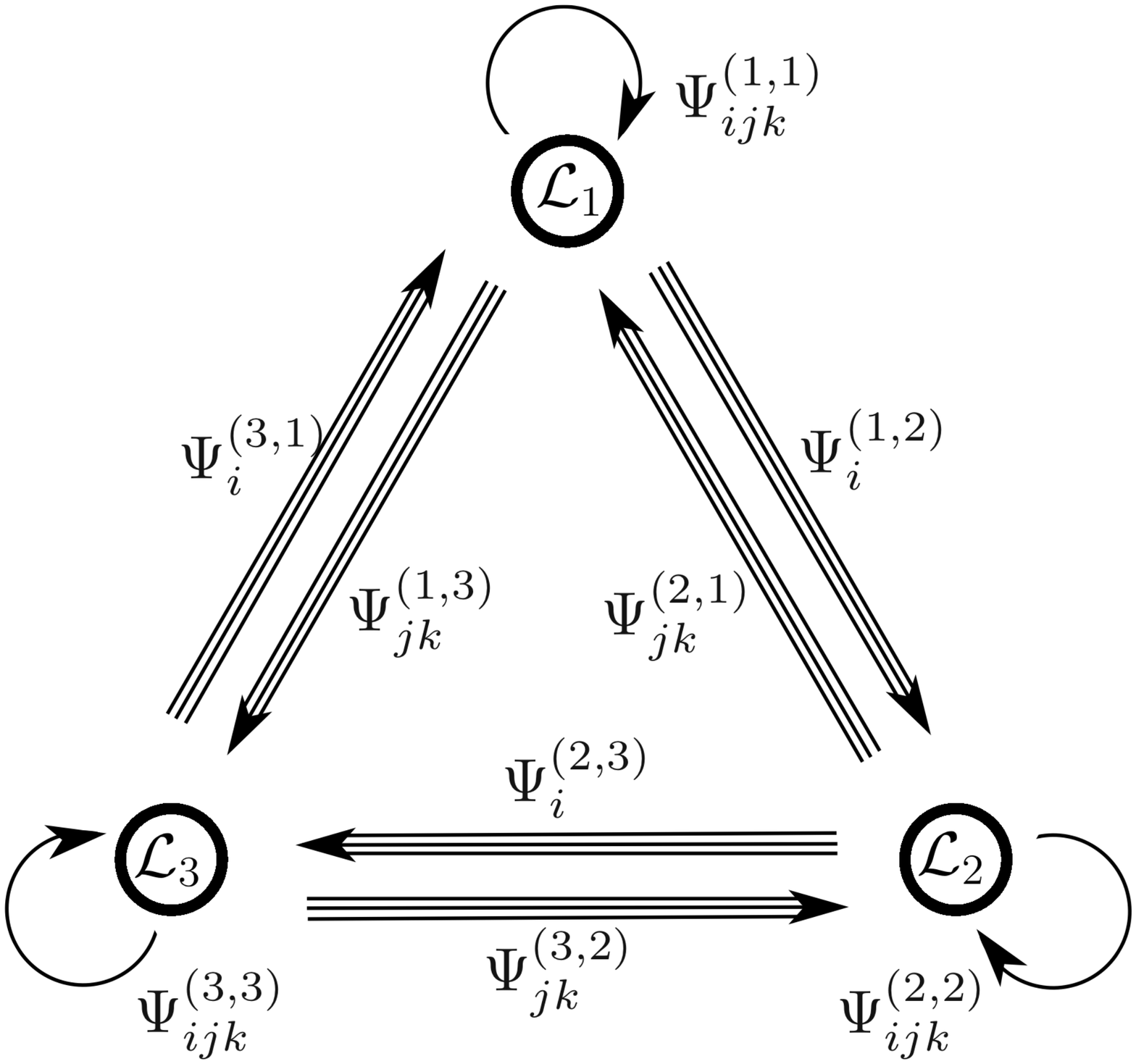}
\caption{Quiver diagram depicting the spectrum of boundary preserving and boundary changing operators.
}
\ll{fig:T2quiver}
\end{center}
\end{figure}

So far, we have considered the topological $B$-model, where the specific brane geometry in question
is encoded  in the canonical matrix factorization \eqz{Qdef} of $W$. All quantities depend on $t$ which is the flat coordinate associated
with the algebraic complex structure deformation of $W$.  As it is usual for LG models, the underlying 
geometry is the one of an orbifold, here  $T_2/\bbZ_3$. 
 
In the present situation we have a self-intersecting brane configuration, so
the operators $\Psi_i$ and $\Psi_{ij}$ are localized on intersections despite not literally being boundary-changing. 
After undoing the orbifold, we obtain an equivariant matrix factorization with 3 different branes, $\cL_1$, $\cL_2$, $\cL_3$, 
 with $RR$ charges $(r,c_1)$ given by  $(2,1)$,$(-1,1)$ and $(-1,2)$ , resp.\ (up to monodromy).  
 The operators localized at the intersections then gain corresponding equivariant labels that govern the selection rules for correlators,
 $\Psi_i\to \Psi_i\lab{A,B}$, $\Psi_{ij}\to \Psi_{ij}\lab{B,A}$ and $\Psi_{ijk}\to \Psi_{ijk}\lab{A,A}$, etc.
 This can be visualized with help of the quiver diagram in {Figure}~\ref{fig:T2quiver}. 
 For background on such equivariant matrix factorizations, see \eg,  \cite{Ashok:2004zb,Govindarajan:2005im}.

We now consider the mirror geometry, where $t$ has the interpretation as a K\"ahler modulus.
The mirror geometry is given by an orbisphere with three punctures, $P_{333}^1\simeq T_2/\bbZ_3$, where again there is just one brane: namely the Seidel special lagrangian which intersects itself three times \cite{Cho:2013kqa}. The fundamental domain is just one-third as compared to the one of the curve, which is why $t$ is implicitly rescaled by a factor of 3.
Undoing the orbifold, the Seidel lagrangian unfolds into three different, pairwise intersecting special lagrangians $\cL_{1,2,3}$ on $T_2$.  
These are shown, on the covering plane, in {Figure}~\ref{fig:instantons}.

$A$-model correlation functions for these brane configurations are well-known \cite{Brunner:2004mt,Herbst:2006nn,Cho:2015}
essentially because for the flat torus the instanton contributions can be just read-off by measuring the areas of polygons. 
These results are particular examples of the general story laid out in refs.\ \cite{Polishchuk:1998db,Polishchuk_homologicalmirror,Polishchuk:2000kx}. 
The correlators have the structure of generalized theta functions, and in our case
the three-point functions are very simple:
\bea
\ll{eq:eqcetacorr}
\langle\, \Psi_1\lab{A_1,A_2}(u_1,u_2)\Psi_{1}\lab{A_2,A_3}(u_2,u_3)\Psi_1\lab{A_3,A_1}(u_3,u_1)\rangle_{D} &=&   \al_1 (\sum u_i, t)\\
\langle\, \Psi_1\lab{A_1,A_2}(u_1,u_2)\Psi_{2}\lab{A_2,A_3}(u_2,u_3)\Psi_3\lab{A_3,A_1}(u_3,u_1)\rangle_{D} &=&   \al_2 (\sum u_i, t)\nn\\
\langle\, \Psi_1\lab{A_1,A_2}(u_1,u_2)\Psi_{3}\lab{A_2,A_3}(u_2,u_3)\Psi_2\lab{A_3,A_1}(u_3,u_1)\rangle_{D} &=&   \al_3 (\sum u_i, t),\nn
\eea
where
\bea
\ll{eq:aTHETAdef}
 \al_\ell (u,t) &=&
e^{2\pi i(\ell/3-1/12)}\,\Theta\Big[{\topa{(1-\ell)/3-1/2}{-1/2} }\,\Big|\,3u,3 t\Big]\ ,  \\
\Theta\Big[{\topa{c_1}{c_2}}\Big|\,n\, u,n t\Big]
&=&
\sum_m q^{n(m+c_1)^2/2} e^{2\pi i(n\, u+c_2)(m+c_1)}\ .
\eea
If the brane moduli are switched off:  $u_\al=0$,  these indeed reproduce our result \eqz{etacorr} due to 
$\al_1 (0, t)=0$ and $\al_2 (0, t)=-\al_3 (0, t)= \eta(t)$. 

This gives then the following $A$-model interpretation of the $\eta$-function 
that we find from the $B$-model:
its first term, $\eta(t)\sim q^{1/24}+....$, measures the area of the smallest triangle as shown in {Figure}~\ref{fig:instantons},
which is $1/24$th of the area of the fundamental domain. The higher powers take the higher wrappings into account.

This result on open-closed Gromov-Witten invariants is by no means new and 
is contained in the previously cited works. However, there the $t$ dependent normalization was put by hand in 
order to fit the known areas of polygons,  and was not computed. In a sense, the explicit mirror map between the two sides
of the isomorphism \eqz{NCVSHS} was missing. Our point was to make predictions for the $A$-model
starting from the $B$-Model, without the need to fix functions by hand. 
This will be important for later applications, \eg\ to Calabi-Yau threefolds, where the $A$-model correlators are not known beforehand.

\begin{figure}[t]  
\begin{center}
\includegraphics[width=11cm]{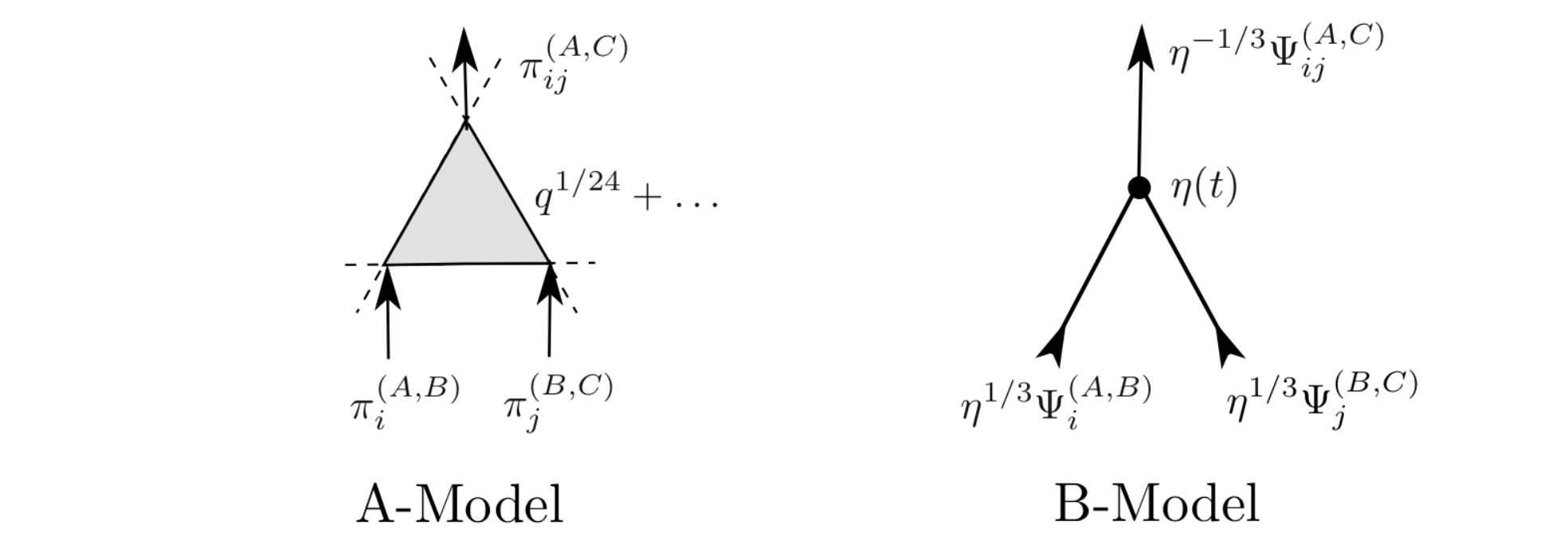}
\caption{Quantum and classical products $m_2$ in the topological $A$- and $B$-models, resp. 
The former features genuine boundary changing
open Gromov-Witten invariants. The latter is entirely
determined by renormalization factors computed by the differential equation \eqz{finalDEQ1}.
}
\ll{fig:Qm2}
\end{center}
\end{figure}

For the four-point amplitude \eqz{bfourpt}, there is a reassuring relation to the works \cite{Cho:2013kqa}.  
The authors used matrix factorizations as well, however on the $A$-model side, which is surprising because in physics, matrix factorizations
appear for $B$-type supersymmetry. What they have done, in a beautiful way, is to interpret the individual
matrix entries of $Q$ as maps which count polygons, and thereby implicitly reproduce the $B$-model matrices 
directly in terms of $A$-model variables. 
Thus $Q$ serves as an $A_\infty$ functor that maps from $Fuk(T_2,\cL_*)$ into $Cat({\rm M\!F},W)$.
At some point they had to impose a ``quantum'' product by hand, and this is precisely the manifestation of the 
structure constant $\eta(t)$ in the product $m_2$  \eqz{m2},
which arises from our $B$-model flatness equations. This is the most basic manifestation of the open string mirror symmetry between quantum products 
in the $A$-model, and classical products in the $B$-model; see {Figure}~\ref{fig:Qm2}.

It was also shown in  \cite{Cho:2013kqa} that by counting triangles one obtains as curvature term of the $A_\infty$ algebra:
\beq
\ll{eq:Wm0}
m_0(s,t)\ =\ \Big[ \varphi(t) {}\sum{s_i}^3- \psi(t) s_1s_2s_3 \Big] \idone\ ,
\eeq
where $\varphi(t)$ and $\psi(t)$ are certain modular functions. By comparison with \eqz{mthrees}, we reproduce them as follows:
 $\varphi\simeq\eta\ft$ and $\psi\simeq\eta\ft\al$, up to constants.
In view of \eqz{bfourpt}, we can explain the results of  \cite{Cho:2013kqa}\ by simply taking $u$-derivatives of \eqz{eqcetacorr}: $\del_u \al_1( u) |_{u=0}\simeq\eta\ft$,
and $\del_u \al_2( u)|_{u=0}=-\del_u \al_3( u)|_{u=0}\simeq\eta\ft\al$. In this way, their results can be directly understood from $B$-model open string correlation functions.

\section{Summary and outlook}

In this paper we made a proposal as to how to compute correlation functions in B-type topological strings
that involve boundary changing operators. This is important because in physics this corresponds to 
computing superpotentials for the largest class of  string backgrounds with $D$-branes, namely ones involving intersecting branes.
This class is infinitely richer than backgrounds without intersecting branes, and extra significance lies also
in the fact that, to our knowledge, such moduli-dependent, boundary changing correlators have 
never been computed (though determined by hand for the elliptic curve).

Summarizing, our main result is a differential equation that is formulated in terms of certain residue pairings.
It generalizes the flatness equations of the bulk theory on the sphere, whose primary component looks
\beq
K\lab0(\nabla_t^{GM}\varphi,*) \equiv\ K\lab0(\del_t\varphi,*)-K\lab1(\phi\varphi,*)\ =\ 0\ .
\eeq
Here $K\lab\ell(*,*)$ are Saito's higher residue pairings, where $K\lab0(*,*)$ is nothing but the topological metric, and 
$K\lab1(*,*)$ has extra insertions of ${d\over dW}$ in it. One can package all the pairings into one quantity, by
summing $K(u)(*,*)=\sum u^\ell K\lab\ell(*,*)$, where $u$ is the degree 2 spectral parameter. Then one can concisely write
\beq
K(u)(\nabla_t^{GM}\varphi,*)\ =\ 0, \ \ \ {\rm with}\ \  \nabla_t ^{GM}= \del_t-{\del_tW\over u}.
\eeq
Our generalization to the boundary theory looks formally similar, except that the pairings are 
formulated in terms of matrix factorizations which underlie $B$-type topological
LG models on the disk. The basic pairing, namely the topological metric, is given
by the Kapustin-Li supertrace residue formula, $K_{KL}\lab0$ given in \eqz{KLori}.
The next higher pairing that we consider, $K_{KL}\lab1$ in \eqz{boundpair}, has
extra insertions of ${d\over dQ}$ in it, 
where $Q$ is the BRST operator that is defined by the factorization $Q^2=W$. 
In terms of these, the flatness equation then looks 
\beq
\langle\langle\nabla_t\Psi\lab{A,B},\Psi\lab{B,A}\rangle\rangle_D = 0,
\eeq
where the ``relative'' boundary-bulk connection is given in  \eqz{finalD}.

The key role at the boundary is played not by $\phi=\del_tW$, but by the boundary counterterm $\ga=\del_tQ$.
This has to do with how stable degenerations of the disk work: there is no direct contact term between $\phi$ and
boundary fields. This also reflects that at the boundary, the relevant
cohomology is given by the one of $Q$, and not of $\cQ_{W} =\bar\del + \iota_{dW}$. Morally speaking,
this suggests to define a boundary connection by $\nabla_t = \idone\del_t-{\del_tQ\over v}$,
where $v$ is a {\it formal} anti-commuting parameter of degree 1 which plays the role of the spectral parameter $u$ in the bulk.

Whether there is more flesh to this than a naive analogy to the bulk, depends on whether one can meaningfully define
higher pairings $K_{KL}\lab\ell$ which would involve more insertions of ${d\over dQ}$.  These should reflect the filtrated Hodge structure
associated with the boundary BRST operator $Q$.
Related questions are how the general definition
of a ``good basis'' in terms of higher residues would look like, in relation to the operator basis that we defined in \eqz{goodbas}.
That should also include a boundary-bulk pairing $K\lab\ell_{bb}(*,*)$ as defined in \eqz{Kbb}, as for $\ell=0$ it is an important ingredient 
of the axiomatic definition of open topological strings 
\cite{LazaroiuTFT,MooreSegal,Costello:2004}.  All-in-all,  conjecturally we would have for a ``good basis'': 
\bea
\ll{eq:allgoodbas}
K\lab0(\varphi_a,\varphi_b)&=& \eta_{ab}\ , \qquad K\lab{\ell>0}(\varphi_a,\varphi_b)\ =\ 0\nn\\
K\lab0_{bb}(\varphi_a,\Psi_\be)&=& \eta_{a \be}\ , \qquad K\lab{\ell>0}_{bb}(\varphi_a,\Psi_\be) =\ 0\ \\
K\lab0_{KL}(\Psi_\al,\Psi_\be)&=& \eta_{\al\be}\ , \qquad K\lab{\ell>0}_{KL}(\Psi_\al,\Psi_\be) =\ 0\ .\nn
\eea
The resolution of these questions would require
a deeper understanding of the underlying mathematics, which is beyond the scope of the paper.  Sorting them out would likely be 
important for the application to higher dimensional Calabi-Yau spaces. This was our main motivation for the present work and
we intend to report on it in the future.

I thank Manfred Herbst, Hans Jockers, Johanna Knapp, Calin Lazaroiu and Johannes Walcher for discussions over the years, and
especially Dmytro Shklyarov for correspondence and comments on the manuscript. This research was also supported in part by the National Science Foundation under Grant No. NSF PHY17-48958.


\bibliography{bibliography}
\bibliographystyle{custom1}


\end{document}